\documentclass[a4paper,11pt]{article}
\pdfoutput=1

\usepackage{jcappub}
\usepackage{aas_macros}
\usepackage{subcaption}
\usepackage{hyperref}
\usepackage{graphicx}
\usepackage{xcolor}
\usepackage{amssymb,amsmath,bm}
\usepackage[utf8]{inputenc}
\usepackage[export]{adjustbox}
\hypersetup{colorlinks=true,citecolor=blue}
\usepackage{multirow}
\usepackage{aas_macros}
\usepackage{subcaption}
\usepackage{amsmath,amssymb,amsfonts}
\usepackage{siunitx}
\usepackage{booktabs}
\usepackage{mathtools}
\usepackage{multirow}
\usepackage{placeins}
\hypersetup{
    colorlinks=true,
    citecolor=blue,
    linkcolor=blue,
    linktoc=page
}

\newcommand{\avg}[1]{\ensuremath{\left\langle \,#1\, \right\rangle}}

\newcommand{\tj}[6]{ \begin{pmatrix}
   #1 & #2 & #3 \\
   #4 & #5 & #6 
  \end{pmatrix}}

\newcommand{\bea}{\begin{eqnarray}}
\newcommand{\eea}{\end{eqnarray}}
\newcommand{\bdm}{\begin{displaymath}}
\newcommand{\edm}{\end{displaymath}}

\def\Mpc{\, h^{-1} \, {\rm Mpc}}

\def\Gpc{\, h^{-1} \, {\rm Gpc}}

\def\kMpc{\, h \, {\rm Mpc}^{-1}}

\def\dk{\frac{\mathrm{d}^3\,k}{(2\pi)^3}\,}

\def\k{{\bf k}}

\newcommand{\eq}[1]{Eq.~(\ref{#1})}
\newcommand{\sect}[1]{Section~\ref{#1}}

\newcommand{\fig}[1]{Figure~\ref{#1}}

\newcommand{\vb}[1]{\mathbf{#1}}
\def\ie{{\em i.e.}~}
\def\eg{{{\em e.g.}~}}

\def\fnl{f_{\mathrm{NL}}^{\mathrm{loc}}}
\def\dq{\int \frac{\mathrm{d}^3q}{(2\pi)^3}}
\def\dpp{\int \frac{\mathrm{d}^3p}{(2\pi)^3}}
\def\dk{\int \frac{\mathrm{d}^3k}{(2\pi)^3}}
\def\eik{e^{i\vb{k}\cdot\vb{x}}}
\def\dlP{\frac{\mathrm{d} \log P}{\mathrm{d}\log k}}
\def\dlM{\frac{\mathrm{d} \log \mathcal{M}}{\mathrm{d}\log k}}

\title{Local Primordial Non-Gaussianities and Super-Sample Variance}
\author[a]{Emanuele Castorina}
\author[b]{and Azadeh Moradinezhad Dizgah}

\affiliation[a]{Theoretical Physics Department, CERN, 1211 Geneva 23, Switzerland}
\affiliation[b]{D\'epartement de Physique Th\'eorique,
Universit\'e de Gen\`eve, 24 quai Ernest Ansermet, \qquad 1211 Gen\`eva 4, Switzerland}
\emailAdd{\textcolor{blue}{emanuele.castorina@cern.ch, Azadeh.MoradinezhadDizgah@unige.ch}}

\abstract{
Fluctuations with wavelengths larger than the volume of a galaxy survey affect the measurement of the galaxy power spectrum within the survey itself. In the presence of local Primordial Non- Gaussianities (PNG), in addition to super-sample matter density and tidal fluctuations, the large-scale gravitational potential also induces a modulation of the observed power spectrum. In this work we investigate this modulation by computing for the first time the response of the redshift-space galaxy power spectrum to the presence of a long wavelength gravitational potential, fully accounting for the stochastic contributions. For biased tracers new response functions arise due to couplings between the small-scale fluctuations in the density, velocity and gravitational fields, the latter through scale dependent bias operators, and the large-scale gravitational potential. We study the impact of the super-sample modes on the measurement of the amplitude of the primordial bispectrum of the local-shape, $\fnl$, accounting for modulations of both the signal and the covariance of the galaxy power spectrum by the long modes.  Considering DESI-like survey specifications, we show that in most cases super-sample modes cause little or no degradation of the constraints, and could actually reduce the errorbars on $\fnl$ by (10 - 30)\%, if external information on the bias parameters is available. 
}

\begin{document}

\maketitle

\vspace{.5cm}
\section{Introduction}

Statistical homogeneity and isotropy of the cosmological fields are two key assumptions in cosmology. They imply that the two-point correlation function, and in general any $n$-point function, is \emph{statistically} invariant under translation and rotation. However, observational effects can break these assumptions. A well known example is the redshift-space distortions, \ie the contribution to the measured redshift of an object of its peculiar velocity \cite{Kaiser:1987qv}, which partially breaks isotropy.
In particular since homogeneity and isotropy are properties of the statistical correlators of the fields and not of fields themselves, they will hold only if we can take the proper ensemble average over the full observable Universe. 

In practice, in galaxy surveys we only observe a finite volume of our past light-cone, and we cannot determine a priori whether the super-survey modes (fluctuations larger than the volume of a given survey) correspond to the mean cosmological value or if they take a non zero finite value. This fact per se does not automatically imply that the measured $n$-point functions do not correspond to the cosmological averages. But it is also essential that the fields  evolve non-linearly and structure form. Only in this case small-scale modes inside the survey can couple to the long-wavelength mode of the size of the survey or larger is possible. 

The effect of the isotropic part of the long modes on the power spectrum has been studied extensively \cite{Hamilton:2005dx,Sirko:2005uz,Hu:2002we,dePutter:2011ah,Baldauf:2011bh,Takada:2013wfa,Wagner:2015gva,Li:2014jra,Martino:2009dm,Seljak:2012tp,Li:2014sga,Chiang:2014oga,Feix:2013hha,Barreira:2017sqa,Schaan:2014cpa,Bertolini:2016hxg}. For instance, it can be used to measure the bias of dark matter halos or of the Lyman-alpha forest \cite{McDonald:2001fe,Li:2015jsz,Baldauf:2015vio,Lazeyras:2015lgp,Modi2016}. Recently, the effects of the tidal part of the super-sample mode has also been investigated \cite{Akitsu:2016leq,Akitsu:2017syq,Chiang:2018mau,Li:2017qgh,Stucker:2020fhk,Schmidt:2018hbj,Masaki:2020drx}. 
It has been shown that in redshift-space both the mean and the tidal part of the long modes contribute to further breaking rotational invariance, causing shifts in the inferred cosmological parameters \cite{Li:2017qgh}. Due to their stochastic nature, super-sample modes have been traditionally considered as an additional source of noise in the covariance matrix of the power spectrum and bispectrum \cite{Hamilton:2005dx,Hu:2002we,Mohammed:2014lja,Chan:2017fiv,Wadekar2019}, hence the name Super-Sample Covariance (SSC). Alternatively one can consider long modes as extra signal \cite{Li:2014sga} and marginalize over their amplitude.

In a Universe with Gaussian initial conditions (ICs), the response of the power spectrum depends only on the mean density and tidal super-sample modes. But the Gaussianity of the ICs is an assumption of the baseline cosmological model, the breaking thereof could point towards new physics in the very early Universe. The simplest models of inflation with only one degree of freedom, \ie the inflaton, with a canonical kinetic term and starting from a Bunch-Davies vacuum state, predict a nearly Gaussian distribution of primordial fluctuations. Stringent constraints on, or detection of PNG would allow distinguishing between different inflation models generating the seed of the observed structure \cite{Alvarez:2014}. The next generation of CMB experiments \cite{Abazajian:2016,Munchmeyer:2018eey} and galaxy surveys \cite{Tellarini:2016sgp,Karagiannis,MoradinezhadDizgah:2018pfo,MoradinezhadDizgah:2018ssw,Gualdi:2020ymf}, as well as potential intensity mapping surveys with various spectral lines \cite{Camera:2013kpa,MoradinezhadDizgah:2018lac,Karagiannis:2019jjx} offer promising possibilities of constraining several shapes of PNG, beyond the current best constraints by Planck satellite \cite{PlanckFNL}. Among various shapes, the local PNG is of particular interest, both theoretically and also because of its observational prospect. This type of PNG can be phenomenologically parameterized by adding a quadratic contribution to the primordial gravitational field $\phi = \varphi_G + \fnl (\varphi_G^2-\avg{\varphi_G^2})$, with $\varphi_G$ being a Gaussian field and $\fnl$ the amplitude of the non Gaussian contribution. In single-field models of inflation, the primordial bispectrum of the local-shape is expected to be nearly zero, independent of the details of the model \cite{Maldacena:2003,Creminelli:2004}. Therefore, a detection of local PNG is considered a smoking gun of multi-field models of inflation. In addition to imprints on the 3-point statistics of the LSS \cite{JeongKomatsu,Sefusatti2009, Baldauf11,Sefusatti:2011gt}, local PNG also leave a unique imprint on the 2-point statistics of biased tracers on large scales \cite{Dalal:2008,Slosar:2008,Matarrese:2008,Afshordi:2008ru}. This signature, referred to as scale-dependant bias has been used to constrain local PNG from current generation of galaxy surveys \cite{Leistedt:2014, Giannantonio:2014b,Castorina2019}. While the errorbars, $\sigma({\fnl}) \simeq 25$, are still larger than the ones from CMB data $\sigma(\fnl)\simeq 5$, they are expected to dramatically improve for the upcoming galaxy surveys \cite{Castorina2019,DESI:2016,Dore:2014}. Taking advantage of cosmic variance cancellations techniques can play an important role in reaching the target sensitivity of $\sigma(\fnl) \leq 1$, using measurements of galaxy power spectrum only \cite{Seljak:2009,Castorina_zerobias,Ginzburg:2019xsj}. 

In this paper we study the effect of the super-sample gravitational potential on the galaxy power spectrum. This type of long-short modes correlation is present only if the ICs are non Gaussian. By focusing on the local case, our work expands on Refs. \cite{Chiang:2017vuk,dePutter:2018} by considering redshift-space distortions and accounting for the correlation between the small-scale tidal fields and the super-sample gravitational potential. 

The rest of the paper is organized as follows. After setting up the notation in the rest of this section and outlining the survey specifications we use, in \sect{sec:rspace}, we present the calculation of the response of the real-space galaxy power spectrum to the presence of super-sample modes, while in \sect{sec:zspace}, we extend the computation to redshift-space. We then discuss the implication of our results for a determination of $\fnl$ from data using Fisher matrix approach in \sect{sec:fisher}, and present the summary and future outlook in \sect{sec:conclude}.

\subsection{Notation}
In a survey of volume $V_s$ and typical size $L_s\simeq V_s^{1/3}$, the main observable is the product of the underlying galaxy density field, $\delta_g(\vb{x})$, with the survey window function, $W(\vb{x})$,
\begin{align}
    \hat{\delta}_g(\vb{x},z) = \delta_g(\vb{x},z) W(\vb{x})\,.
\end{align}
In this work variables with a $\hat{()}$ indicate quantities estimated/measured within the survey.
Over the full, but still finite, volume $V_s$, the mean value of the dark matter overdensity field doesn't have to be zero, \ie the cosmological mean, but it is instead given by
\begin{align}
    \Delta_0 = \dpp \delta(\vb{p},z)W(-\vb{p})\, ,
\end{align}
where $\delta(\vb{p},z)$ and $W(-\vb{p})$ are the Fourier Transform of the density field and window function respectively. Note that we have drop the explicit redshift dependence of the long modes. When no confusion arises we will use the same symbol for a variable and its Fourier Transform.
We are also interested in the mean tidal field in the survey $\tau_{ij}$,
\begin{align}
    \tau_{ij} = \dpp \left( p_i p_j -\frac{1}{3}\delta_{ij}^K\right) \delta(\vb{p},z)W(-\vb{p})\,
\end{align}
and more precisely in its projection along a certain direction $\hat{n}$. We follow \cite{Li:2017qgh} and define the isotropic ($L=0$) and tidal part ($L=2$) of the long wavelength modes as
\begin{align}
    \Delta_L(\hat{n}) = \dpp \delta(\vb{p},z)W(-\vb{p}) \mathcal{L}_L(\hat{n}\cdot\hat{p})\, ,
    \label{eq:Delta}
\end{align}
where $\mathcal{L}_L$ are Legendre polynomials.

For simplicity, in this work we assume the window function is spherically symmetric and normalized to unity, \eg a spherical top-hat, such that the variance of the long mode reads
\begin{align}
    \sigma_L^2 = \frac{1}{2L+1} \int \frac{\mathrm{d} k}{2\pi^2} \, k^2 P(k,z) W^2(k)\, ,
\end{align}
in terms of the linear dark matter power spectrum $P(k,z)$.
The mean value of the DM density field in the survey volume, $\Delta_0$, is thus a number drawn from a Gaussian with mean zero and variance $\sigma_0^2$, and similarly for the tidal field.
The same arguments apply to estimate the value of the long wavelength gravitational potential $\phi_0$, with $P(k,z)$ replaced by $P_\phi(k) = P(k,z) \mathcal{M}^{-2}(k,z)$, with
\begin{align}
 \mathcal{M} (k,z) \equiv\frac{2 c^2 k^2 T(k) D(z)}{3 \Omega_m H_0^2}\, ,
\end{align}
where $T(k)$ is the linear transfer function, $c$ is the speed of light, $D(z)$ is
the linear growth factor normalized to $(1+z)^{-1}$ in the matter-dominated
era, $\Omega_m$ is the matter density parameter
at $z=0$, and $H_0$ is the present-day
Hubble parameter.
To compute the variance of $\phi_0$ defined as
\begin{align}
    \sigma_\phi = \int \frac{\mathrm{d} k}{2\pi^2} \, k^2 P_\phi(k) W^2(k)\, ,
\end{align}
we need to impose a cut-off at low-$k$, which we choose to be the present day horizon\footnote{We checked that changing the IR cutoff to the present day Hubble scale $H_0^{-1}$ does not qualitatively change any of the results.}.
In this work we assume a Planck+BAO fiducial cosmology \cite{PlanckCosmo}.

\subsection{Survey specification and fiducial galaxy biases}\label{sec:specs}
In this work we consider two galaxy samples to show the effects of the super-sample modes, one at $z\simeq1$ with linear bias of $b_1=1.35$ and one at $z=2.5$ with linear bias of $b_1=4$. Loosely speaking they could be identified with the ELG and QSO sample of DESI \cite{DESI:2016}. The value of the shot-noise for the two sample is $N[\Mpc)^3] = \{ 3\times 10^3, \,10^5 \} $, for the low- and high-$z$ sample respectively. Other important parameters in computing the Fisher forecast are the volume of the survey, $V_s$, which sets the largest available scale $k_{\rm min} = 2\pi/V_s^{(1/3)}$, and the largest wavenumber included in the analysis $k_{\rm max}$. For $V_s$ we take the volume corresponding to roughly the entire ELG or QSO sample, $V_{ELG}\simeq (3.5 \Gpc)^3$ and $V_{QSO}\simeq (5.5 \Gpc)^3$. We will show results for different choices of $k_{\rm max}$. 

Let us also outline the choices of the values of the galaxy biases used in describing the galaxy overdensity field given in Eq. \eqref{eq:bias}, that we use throughout the paper, both in computing the response functions and in the Fisher forecasts. For the second-order in density bias $b_2$, we use the fitting formula presented in Ref. \cite{Lazeyras:2015lgp} to relate it to linear bias, while for the second-order tidal bias $b_{s^2}$ we take the co-evolution prediction $b_{s^2} = -2/7(b-1)$. We use the Peak-Background-Split to fix the fiducial value of the Non-Gaussian biases (see \cite{BiagettiRev} for a review)
\begin{align}\label{eq:png_bias}
    b_\phi = 2 \delta_c (b_1-1)\;,\;b_{\phi\delta} = 2 [\delta_c( b_2 + 13/21 (b_1 - 1)) - b_1 + 1]\;.
\end{align}

\section{Responses in real-space}
\label{sec:rspace}
In the presence of local-shape primordial non-Gaussianity, in addition to matter density field and tidal tensor, the galaxy over-density also depends on gravitational potential. Expanding in terms of renormalized operators, the galaxy overdensity up to second order in perturbation theory is given by \cite{GiannantonioPorciani,Assassi:2015fma,Desjacques:2016} 
\begin{align}
    \delta_g (\vb{x},z) &= b_1 \delta (\vb{x},z) + \frac{1}{2}b_2 \delta^2(\vb{x},z) + b_{s^2} s^2(\vb{x},z) \nonumber \\
    &+\fnl[ b_\phi \phi(\vb{q},z) + b_{\phi \delta} \delta(\vb{x},z) \phi(\vb{q},z)] \nonumber  \\[0.7ex]
    &+ \epsilon(\vb{x}) + \epsilon_\delta(\vb{x}) \delta(\vb{x}) + \fnl\epsilon_{\phi}(\vb{x}) \phi(\vb{q},z) \, ,
    \label{eq:bias}
\end{align}
where we have only kept the terms linear in $\fnl$ and neglected the contributions from higher-order derivative operators. To avoid clutter we have dropped the explicit redshift-dependence of the bias parameters. Here $\phi$ is the primordial gravitational potential, and $\vb{q} = \vb{x} - \vb{\Psi}(\vb{q})$ is the Lagrangian coordinate, which  at leading order is related to the linear density field by $\delta_L = -\nabla \vb{\Psi}$. The second order field $s^2$ corresponds to the traceless part of the shear field and is defined as \begin{align}
   s^2(\vb{x},z) = \dk \,\eik\,\dq \left(\left[\frac{\vb{q}\cdot (\vb{k}-\vb{q})}{|\vb{q}| |\vb{k}-\vb{q}|}\right]^2-\frac{1}{3}\right) \delta(\vb{q})\delta(\vb{k}-\vb{q})\, .
\end{align}
We include stochastic terms in the last line \eq{eq:bias}, which we assume to be of Poisson origin. They will play the role of the noise in the Fisher analysis.

The linear galaxy power spectrum according to this bias model reads
\begin{align}
    P_g(k,z) = \left[b_1^2 + 2 \fnl b_1 b_\phi \mathcal{M}^{-1}(k,z)\right] P(k,z) + N \, ,
\end{align}
where $N\equiv \avg{\epsilon \epsilon} = 1/\bar{n}$ is the Poissonian shot-noise, and we have only kept terms linear in $f_{\rm NL}^{\rm loc}$. The measured power spectrum $\hat{P}_g(k)$ in a survey of  finite volume $V_s$, however, is the average of the galaxy fluctuations for a fixed realization of the long modes; therefore, does not necessarily correspond to the cosmological average power spectrum $P_g(k)$. The power spectrum of short-scale modes within the survey volume in the presence of super-survey modes can be schematically written as
\begin{align}
    \hat{P}_g(k,z)  \equiv \avg{\delta_g(\vb{k},z)\delta_g^*(\vb{k},z) | \Delta_0, \phi_0} = P_{g}(k,z) + R_\Delta (k,z) \Delta_0 + R_\phi(k,z) \phi_0\,,
\end{align}
% \begin{align}
%     \hat{P}_g(k)  \equiv \avg{\delta_g(\vb{k})\delta_g^*(\vb{k})}_{\Delta_0, \phi_0} =\avg{\delta_g(\vb{k})\delta_g^*(\vb{k}) | \Delta_0, \phi_0} = P_{g}(k) + R_\Delta (k,z) \Delta_0 + R_\phi(k,z) \phi_0\,.
% \end{align}
where, the functions $R_\Delta$ and $R_\phi$ are called response functions. Notice the long wavelenghts modes $\Delta_0$ and $\phi_0$ are in configuration space, while the galaxy perturbations are in Fourier space. The modulation of the measured power spectrum by the long wavelength modes can be seen as an extra term in the power spectrum covariance, hence the name super-sample variance. At the same time we can think of it as extra signal, with the amplitude of $\Delta_0$ and $\phi_0$  the two new free parameters one has to marginalize over when constraining cosmological parameters.

We can compute the response functions from the squeezed limit of the bispectrum $B_g(\vb{p},\vb{k}_1,\vb{k}_2,z)$,  where one mode is much longer than the other two, $p\ll k_1\simeq k_2 = k$ \cite{Wagner:2014aka,Chiang:2014oga,Li:2014jra,Barreira:2017sqa}. This configuration captures the correlation between one large-scale mode and two small-scale ones that we are interested in. One then has to average over the angular part of the super-sample mode $\hat{p}$ since it is unknown.
The angle-averaged squeezed-limit of galaxy bispectrum is related to response functions as
\begin{align}
  \int \frac{\mathrm d \Omega_{\hat{p}}}{4 \pi}  B_g^{\rm sq}(\vb{p},\vb{k}_1,\vb{k}_2,z) &\equiv \lim_{p\ll k_1,k_2} \int \frac{\mathrm d \Omega_{\hat{p}}}{4 \pi}  B_g(\vb{p},\vb{k}_1,\vb{k}_2,z) \nonumber \\[0.7ex]
  &= b_1 R_\Delta(k,z) P(p,z)   +  [b_1 R_\phi(k,z) + \fnl b_\phi R_\Delta(k,z)]P_{\phi \delta}(p,z)\, .
    \label{eq:real_space}
\end{align}
The explicit expression of the galaxy bispectrum at tree level, which we use to derive the response functions, and includes the contributions from primordial non-Gaussianity and gravitational evolution is given in Appendix \ref{app:bis_tree}.  More details on the derivation can be found in Refs. \cite{Li:2017qgh, Chiang:2017jnm,dePutter:2018jqk}.

The response function $R_\Delta$ originates from the coupling between the large-scale density field and the small-scale modes, either in density or gravitational potential. It contains a Gaussian piece, due to nonlinear evolution, and a non-Gaussian one of primordial origin,
\begin{align}
    R_\Delta (k,z) &= 
    \left[\frac{47}{21}b_1^2+ 2 b_1 b_2 - \frac{b_1^2}{3}\dlP \right] P(k,z) +\frac{b_1 }{\bar{n}} \nonumber \\
    & + \fnl\left[\frac{26}{21}b_1 b_\phi+2 b_1 b_{\phi\delta} + 2 b_2 b_\phi -\frac{2}{3} b_1 b_\phi\left( \dlP- \dlM\right) \right] P_{\phi \delta}(k,z) \nonumber \\[0.7ex]
    & \equiv R^{\rm G}_\Delta(k,z) +\fnl R^{\rm NG}_\Delta(k,z) \,.
\end{align}
Notice $R^{\rm NG}_\Delta(k,z)$ contains only terms proportional to PNG bias parameters, \ie they correspond to couplings between $\Delta_0$ and the short wavelength $\phi$. 
In other words $R^{\rm NG}_\Delta(k,z)=0$ for the response of the dark matter power spectrum to the long mode density field even for non Gaussian initial conditions. The presence of a long mode changes both the expansion history and the growth of dark matter fluctuations \cite{Martino:2009dm}.
The terms proportional to $P(k)$ or $P_{\phi \delta}$ are often called growth terms \cite{Li:2014sga}, and arise because positive (negative) amplitude of the long modes enhances (reduces)  the growth of structure. The terms proportional to derivatives of power spectra instead show the effect of the long modes on the expansion history and are usually called dilation terms \cite{Li:2014sga}. The shot-noise term comes from the following contribution of the stochastic operators to the Bispectrum
\begin{align}
    B^{\rm sq}(\vb{p},\vb{k}_1,\vb{k}_2,z) &\supset b_1 \avg{ \delta (\vb{p},z) \left[\epsilon (\vb{k}_1) + \frac{1}{2} \dq \epsilon_\delta (\vb{q}) \delta(\vb{k_1}-\vb{q})\right]\epsilon(\vb{k}_2)} + \vb{k}_1 \xleftrightarrow{}\vb{k}_2 \notag\\[0.7ex]
    &= b_1 \avg{\epsilon_\delta \epsilon} \avg{ \delta(\vb{p},z)\delta(\vb{k}_2,z)} +  \vb{k}_1 \xleftrightarrow{}\vb{k}_2 \nonumber \\[0.7ex]
    &=2 b_1 P_{\epsilon \epsilon_\delta} P(p,z) = \frac{ b_1^2 P(p,z)}{\bar n}\,.
    \label{eq:noise}
\end{align}

The other response function $R_\phi$ contains couplings between $\phi_0$ and the small-scales fields; therefore, it is identically zero in the absence of PNG
\begin{align}
    R_\phi(k,z) = 2 \fnl \left[(2 b_1^2+ b_1 b_{\phi \delta})P(k,z)  +\frac{ b_\phi}{2\bar{n}}\right]\,.
    \label{eq:rphi}
\end{align}
The first term in the above equations comes from the primordial Bispectrum and it would be there even for dark matter, while the second one is present only for biased tracers.
The response to $\phi_0$ contains only growth terms, since the response to $\fnl$ is locally equivalent to rescaling of the amplitude of the fluctuations \cite{Slosar:2008}.

\begin{figure}
    \centering
    \includegraphics[width=\textwidth]{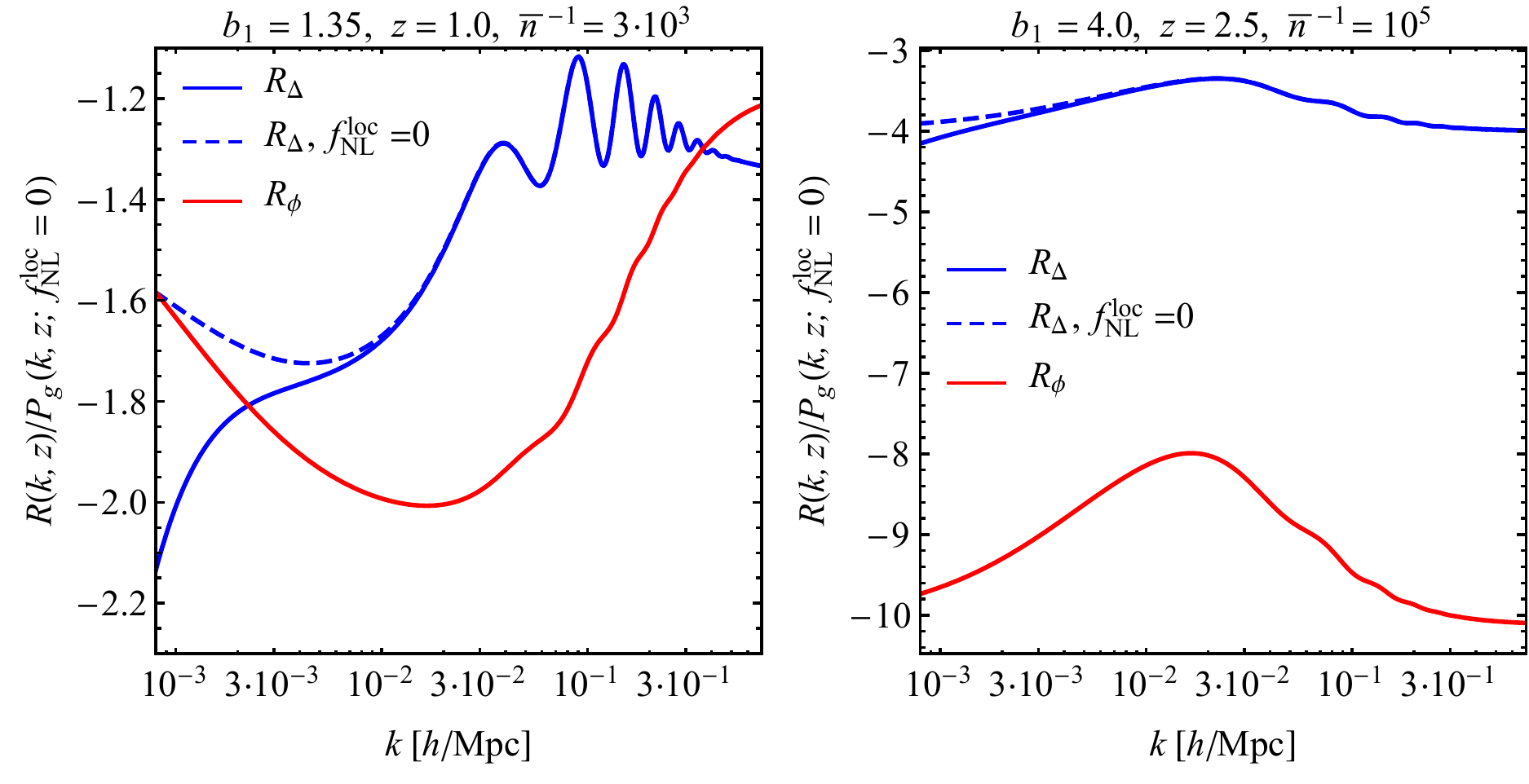}
    \caption{Real Space response functions for two different configurations: a DESI-ELG like on the left, and a DESI-QSO like on the right.}
    \label{fig:real}
\end{figure}

In redshift surveys the galaxy overdensity is usually estimated by computing the mean number of objects $\hat{\bar{n}}_g$ within the survey itself. The latter is also affected by the presence of long wavelength galaxy fluctuations $\Delta_g$,
\begin{align}
    \hat{\bar{n}}_g = \bar{n}_g(1+\Delta_{g})=\bar{n}_g(1+b_1\Delta_0 +  \fnl b_\phi \phi_0)\,.
\end{align}
For a power spectrum normalized by $\hat{\bar{n}}_g^{-2}$ the response functions become
\begin{align}
    R_\Delta(k,z) \longrightarrow R_\Delta(k,z)-2b_1P_g(k,z)\;,\;   R_\phi(k,z) \longrightarrow R_\phi(k,z)-2 \fnl b_\phi P_g(k,z)\,.
    \label{eq:mean}
\end{align}
Note that in the above equation, when computing $R_\phi$, we drop the term proportional to $f_{\rm NL}^{\rm loc}$ in $P_g(k,z)$ since their contribution to $R_\phi$ would be quadratic in $f_{\rm NL}^{\rm loc}$. In the literature the response functions that include the piece arising from the normalization of the density field are usually called the \emph{local} ones, whereas the \emph{global} ones do not have this extra term.
Note that if one chooses instead to normalize the power spectrum by $\bar{n}_g^{-1}$, as usually done in galaxy survey with the FKP estimator \cite{Wadekar2019}, then the one has to drop the factor of 2 in the second term in \eq{eq:mean}. For galaxy surveys the local responses are the relevant ones, so we will stick to them in the rest of this work.

Before discussing the shape of the response functions, let use make two additional notes regarding the shot-noise contributions. First, the factor of $P_g(k,z)$ in the above equations includes the shot noise contribution to the galaxy power spectrum. This term partially cancels  with the shot noise contribution to the squeezed limit of the Bispectrum. The cancellation will be exact for a FKP estimator. It is important to notice the cancellation holds only for Poissonian shot-noise. Compared to previous work, our derivation of the shot noise contribution to the super-sample signal using the squeezed limit of the Bispectrum highlights the physical difference between the normalization of the power spectrum and the terms in \eq{eq:noise}. Second, in \eq{eq:rphi} and \eq{eq:mean} we see that the shot noise induces new PNG terms. One could be tempted to consider it an extra signal, but it is easy to see this contribution just changes the value of the true shot noise which is always marginalized over as a free parameter.
%For this reason when plotting the response functions we will remove the constant, in $k$, shot noise contribution to $R(k,z)$'s.
It is however important to keep noise terms in the super-sample signal in the forecast analysis, discussed in \sect{sec:fisher}, as they increase the variance of the power spectrum.

\fig{fig:real} shows the real-space response functions for low- and high-redshift samples for a DESI-like survey described in Section \ref{sec:specs}. We have set the value of $\fnl=1$, assumed the second-order bias $b_2$ as a function of $b_1$ according the fit presented in Ref. \cite{Lazeyras:2015lgp}, and set the value of $b_{s^2}$ using the coevolution prediction \cite{Chan:2012jj,Baldauf:2012hs}. The effect of PNG on $R_\Delta$ can be seen on large scales where the difference between the blue and the dashed blue lines is manifest. This was expected since the non-Gaussian part of the response to $\Delta_0$ is proportional to $P_{\phi \delta}\ll P(k)$ at high-$k$. Both growth and dilation terms contribute to $R_\Delta$, as one can notice from the oscillations around the BAO scale. As discussed above the response $R_\phi$, shown in red, does not contain dilation terms, hence no large wiggles are present.  Both responses are negative because the dominant contribution is coming from the rescaling of the mean in \eq{eq:mean}.

\section{Responses in redshift-space}
\label{sec:zspace}
The redshift-space response functions are also straightforward to calculate. The main difference with respect to the real-space calculation is that RSD break isotropy of space; therefore, one expects a different response to the isotropic and shear part of the long modes.
We start from the expression for the second order galaxy overdensity field in redshift-space, $\delta_g^s(\vb{k})$,
\begin{align}
    \delta_g^s(\vb{k},z) &= \delta_g(\vb{k},z) + f \mu^2_k \theta(\vb{k},z)- \frac{f \mu_k k}{2} \dq [\delta_g(\vb{q},z)+ f \mu^2_q \theta(\vb{q},z)] \notag \\
    &\times \frac{(\vb{k}-\vb{q})\cdot \hat{n}}{|(\vb{k}-\vb{q})|^2} \theta (\vb{k}-\vb{q},z)+ \vb{q} \xleftrightarrow{} (\vb{k}-\vb{q})\, ,
\end{align}
where $f$ is the linear growth rate, $\theta(\vb{k},z)$ is the divergence of the velocity field, and $\hat{n}$ is the line of sight (LOS) direction. We work in the plane parallel-limit and neglect wide angle/curved sky corrections to the above formula \cite{CastorinaWhite2018a,CastorinaWhite2018b,BeutlerCastorina}. In the presence of local-shape PNG, the galaxy power spectrum at tree-level, including the linear RSD \cite{Kaiser:1987qv}(Kaiser term), is given by
\begin{align}
    P_g^s(k,\mu,z) =  \left[(b_1 + f\mu^2)^2 + 2 \fnl (b_1 + f \mu^2) b_\phi \mathcal{M}^{-1}(k,z)\right] P(k,z) + N\, .
\end{align}
In the squeezed limit the bispectrum between one long-wavelength real space galaxy mode, $\delta_g(\vb{p})$, and two redshift-space small-scales modes can be written in the following way
\begin{align}
   \lim_{p\ll k_1,k_2}   B_g^s(\vb{p},\vb{k}_1,\vb{k}_2,z) &= \lim_{p\ll k_1,k_2}   \avg{\delta_g(\vb{p},z) \delta_g^s(\vb{k}_1,z) \delta_g^s(\vb{k}_2,z)} \notag \\ &=\sum_{\ell_1, \ell_2} f_{\ell_1,\ell_2} (p,k,\mu_k,z) \mathcal{L}_{\ell_1}(\nu)\mathcal{L}_{\ell_2}(\mu_p)\, ,
\end{align}
where $\mu_k = \hat{k}\cdot \hat{n}$, $\nu \equiv \hat{k}\cdot \hat{p}$ and $\mu_p \equiv \hat{p}\cdot \hat{n}$. 
\begin{figure}[h]
    \centering
    \includegraphics[width=\textwidth]{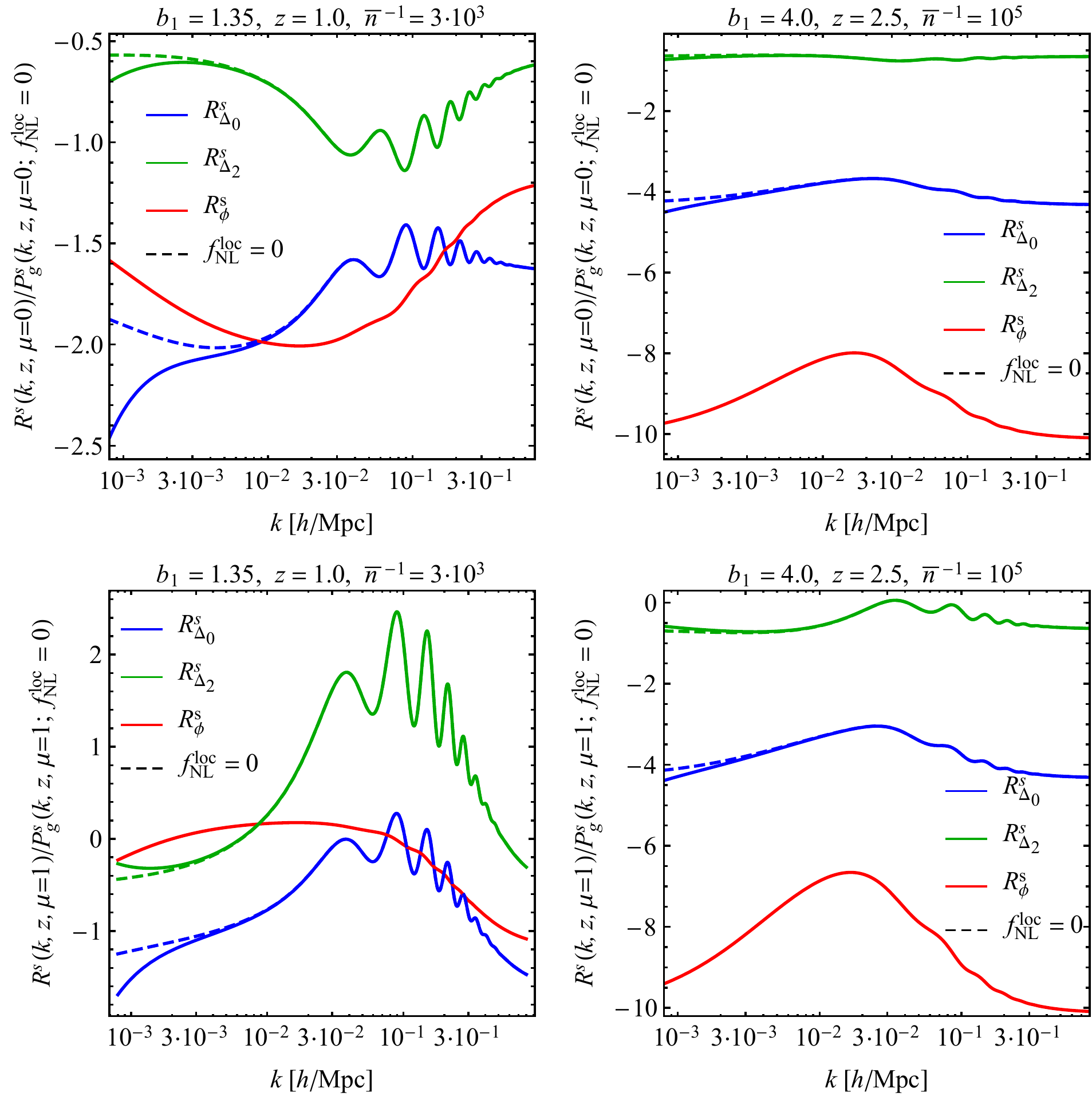}
    \caption{Redshift-space response functions for two different configurations: a DESI-ELG like on the left, and a DESI-QSO like on the right.}
    \label{fig:zspace}
\end{figure}

Similar to the real-space computation, to compute the responses we just have to average the squeezed Bispectrum with the appropriate weight according to the definition of the long modes in \eq{eq:Delta},

\begin{align}
(2\ell+1)  \lim_{p\ll k_1,k_2}  &\int  \frac{\mathrm{d}\Omega_{\hat{p}}}{4 \pi}  B_g(\vb{p},\vb{k}_1,\vb{k}_2,z) \mathcal{L}_{\ell}(\mu_p) \notag \\
&= (2\ell+1)\int \frac{\mathrm{d} \Omega_{\hat{p}}}{4 \pi}  \sum_{\ell_1, \ell_2} f_{\ell_1,\ell_2} (p,k,\mu_k,z) \mathcal{L}_{\ell_1}(\nu)\mathcal{L}_{\ell_2}(\mu_p) \mathcal{L}_{\ell}(\mu_p) \notag \\
& = (2\ell+1)\int \frac{\mathrm{d} \Omega_{\hat{p}}}{4 \pi}  \sum_{\ell_1, \ell_2} f_{\ell_1,\ell_2} (p,k,\mu_k,z) \mathcal{L}_{\ell_1}(\nu) \sum_L \tj{\ell_2}{\ell}{ L}{0}{0}{0}^2(2L+1)\mathcal{L}_{L}(\mu_p) \notag \\
& = (2\ell+1) \sum_{\ell_1, \ell_2} f_{\ell_1,\ell_2}(p,k,\mu_k,z) \tj{\ell}{\ell_1}{ \ell_2}{0}{0}{0}^2 \mathcal{L}_{\ell_1}(\mu_k)\;,
\end{align}
and finally read off the terms proportional to power spectra of the super-sample modes as in \eq{eq:real_space}.
We schematically write 
\begin{align}
   \hat{P}_{g}^s(k,\mu_k,z) &\equiv \avg{\delta_g^s(\vb{k},z)\delta_g^{s,*}(\vb{k},z)}_{\Delta_0,\Delta_2,\phi_0} \nonumber \\
   &= P_{g}^s(k,\mu_k,z) + R^s_{\Delta_0}(k,\mu_k,z) \Delta_0+R^s_{\Delta_2}(k,\mu_k,z) \Delta_2 + R^s_{\phi}(k,\mu_k,z) \phi_0\, ,
\end{align}
with the three response functions given by
\begin{align}
    R^s_{\Delta_0}(k,\mu_k,z) &= 
   P(k,z)\left[ -\frac{1}{3}
   \dlP \left(f \mu ^2+1\right) \left(b_1+f \mu ^2\right){}^2 \right. \notag \\   & \left.+\frac{1}{21} \left(b_1+f
   \mu ^2\right) \left(f \mu ^2 \left(42 b_1-7 f+31\right)+7 b_1 f+47 b_1+42 b_2+28 f^2\mu
   ^4\right) \right]  \notag \\
  &+  f_{\text{NL}}^{\text{loc}} P_{\phi \delta}(k,z)\left[\frac{2}{3} \left(\dlM -\dlP\right) b_{\phi } \left(f \mu ^2+1\right)
   \left(b_1+f \mu ^2\right) \right.  
   \notag \\ 
   & \left. + \frac{2}{21} \left(f \mu ^2 \left(b_{\phi } \left(14 f \mu
   ^2+5\right)+21 b_{\phi \delta }\right)+b_1 b_{\phi } \left(7 f \left(3 \mu
   ^2+1\right)+13\right)+21 b_2 b_{\phi }+21 b_1 b_{\phi \delta }\right)\right] \nonumber \\
   & + \frac{(b_1+f/3)}{\bar{n}}\,,
   \label{eq:Rsd0}
\end{align}
 
\begin{align}
 \hspace{-1.3in}  R_\phi^s(k,\mu_k,z)=2 f_{\text{NL}}^{\text{loc}} \left\{ \left(b_1+f \mu ^2\right)  \left[f \mu ^2
   \left(b_{\phi }+2\right)+b_{\phi \delta }+2 b_1\right]P(k,z)  + \frac{b_\phi}{2 \bar{n}}\right\}\,,
   \label{eq:Rsphi}
\end{align}

\begin{align}
     R^s_{\Delta_2}(k,\mu_k,z) & = P(k,z) \left[ \frac{2}{21} \left(b_1+f \mu ^2\right) \left(\mu ^2 \left(12 b_1+42
   b_{s^2}-f (7 f+8)\right)+7 b_1 f-4 b_1-14 b_{s^2} \right. \right. \notag \\ &\left. \left. +4 f (7 f+6) \mu
   ^4\right) -\frac{1}{3} \dlP \left((2 f+3) \mu ^2-1\right) \left(b_1+f \mu
   ^2\right){}^2 \right]
   \notag \\ 
   & + f_{\text{NL}}^{\text{loc}}  P_{\phi \delta}(k,z)\left[\frac{4}{21} b_{\phi } \left(\mu ^2 \left(6 b_1+21
   b_{s^2}-4 f\right)+7 b_1 f-2 b_1-7 b_{s^2}+2 f (7 f+6) \mu ^4\right) \right. \notag \\ & \left. 
   + \frac{2}{3} \left(\dlM-\dlP\right) b_{\phi } \left((2 f+3) \mu ^2-1\right) \left(b_1+f \mu
   ^2\right)\right]  +\frac{2}{3 \bar{n}}f\,.
   \label{eq:Rsd2}
\end{align}
As expected for local PNG there is no response to the tidal part of the long wavelength gravitational potential.
For $\fnl=0$ our expressions agree with Ref. \cite{Li:2017qgh} \footnote{Notice that Ref. \cite{Li:2017qgh} define the responses in terms of $\mathrm{d}\log \mathcal{P} / \mathrm{d} \log k \equiv \mathrm{d}\log k^3 P(k,z)/ \mathrm{d} \log k$.}.
The normalization of the density fluctuations also shifts the redshift-space response functions,
\begin{align}
     &R^s_{\delta,0} \rightarrow R^s_{\delta,0} - 2(b_1 + f/3)P_g(k,\mu_k,z)\, , \nonumber \\
     &R_\phi^s \rightarrow  R_\phi^s- 2 f_{\rm NL}^{\rm loc} b_{\phi}P_g(k,\mu_k,z)\, , \nonumber \\ &R^s_{\delta,2}\rightarrow  R^s_{\delta,2}-4/3 f P_g(k,\mu_k,z)\,.
     \label{eq:meanRSD}
\end{align}
Analogously to the real space calculation the shot-noise contribution to the squeezed limit bispectrum is partially canceled by the change in the mean number density.

The redshift-space response functions are shown in \fig{fig:zspace} for the same two galaxy samples of \fig{fig:real}. The upper panel shows the $\mu=0$ response and the lower one the $\mu=1$ response. As first noted in \cite{Li:2017qgh}, the $\mu=0$ responses do not reduce to the real space ones, as we have already performed an angular average to define them. The real-space responses can be recovered also sending $f\rightarrow0$.
It is worth noticing that in redshift-space the large-scale tidal field couples with PNG bias parameters, \ie $R^s_{\Delta,2}$ contains terms proportional to $\fnl$.
In general the response functions are increasing function of $\mu_k$.
It is straightforward to project the two-dimensional power spectrum into multipoles $P_{\ell}(k)$ but the final expressions are not very illuminating, so we dont show them here.

\section{Super-sample modes and constraints on local PNG}
\label{sec:fisher}
Our next goal is to assess the impact of marginalization over the amplitude of the super-sample modes on $\fnl$ constraints. We use a Fisher Matrix approach for this purpose, assuming a fiducial value of $\fnl=0$. We focus on redshift-space, but results for real space are very similar. The free parameters are $\bm{\theta} \equiv \{b_1,b_2,\gamma_2,N,\fnl,\Delta_0,\Delta_2\}$, where $N$ is the amplitude of the Poissonian shot noise term in the power spectrum. We do not include $\phi_0$ as a free parameter since it always enters  multiplied by $\fnl$ and would therefore make the Fisher matrix singular for $\fnl=0$. As we will see later the value of $\phi_0$ affects the results in some cases, therefore we present the constraint on $\fnl$ for different values of $\phi_0$ in the range $[-5 \sigma_\phi,5 \sigma_\phi]$. The value of $\Delta_L^{\rm fid}$ will change accordingly in the range $[-5 \sigma_L,5 \sigma_L]$, but it has basically no impact on the PNG constraints.

A few points are in order regarding the choices of the varied parameters, the priors and the fiducial values. While we use the fit in Ref. \cite{Lazeyras:2015lgp} to set the fiducial value of $b_2$, it is important to stress that for QSO samples the values of $b_2$ could be very different than the fit to mass selected halos used for the fit. Furthermore, the non-Gaussian bias parameters could deviate from the simple peak-background split prediction shown above \cite{Slosar:2008}. The shot-noise contribution to response function should also be considered as an independent free parameter since it comes from the squeezed limit of the bispectrum, see \eq{eq:noise}. It is however very degenerate with the shot-noise in the galaxy power spectrum, and we will therefore use only a single stochastic free parameter $N$. Finally, when fitting the data from galaxy surveys to constrain PNG, it is a common practice to keep the shape of the power spectrum fixed, \ie the cosmological parameters are given and one marginalizes only over galaxy bias parameters and shot noise. We shall do the same here, which implies a prior on $\Delta_L$ will likely be available. We will show results with and without a prior on the long modes\footnote{The strength of the prior is irrelevant for the final error on $\fnl$.}. When super-sample modes are included we also need to set the fiducial values of $\Delta_L$. We take $\pm \left\{{1,3,5}\right\}$-$\sigma$ values. In order to reduce the noise in the inversion of the Fisher matrix we impose very mild prior of $b_2$ and $b_{s^2}$, $\sigma(b_2)/b_2 = \sigma(b_{s^2})/b_{s^2}=5$.

The Fisher matrix is defined as \cite{Tegmark:1997rp}
\begin{align}
    F_{\alpha \beta} =\sum_{\ell_1,\ell_2,i,j}\frac{\partial \hat{P}_{g,\ell_1}(k_i)}{\partial \theta_\alpha} [C_{\ell_1 \ell_2}(k_i,k_j)]^{-1}\frac{\partial \hat{P}_{g,\ell_2}(k_j)}{\partial \theta_\beta}
\end{align}
where the sum runs over the multipoles of the power spectrum $\ell_1,\ell_2 = 0,2,4,6$ and the binned value of the wavenumber $k_i,k_j$. The binned covariance of the power spectrum multipoles $C_{\ell_1 \ell_2}(k_i,k_j)$ for our fiducial value of $\fnl =0$ has been computed in \cite{Li:2017qgh} and it contains a diagonal piece due to cosmic variance, and diagonal and off-diagonal entries due to super-sample variance (see also \cite{Wadekar2019}). We work at sufficiently small $k$ that the trispectrum contribution to the covariance can be safely ignored \cite{Meiksin:1998mu}. For the results presented in the next Sections the SSC does not play any significant role.

\subsection{High-$z$, high bias sample}
\begin{figure}
    \centering
    \includegraphics[width=1\textwidth]{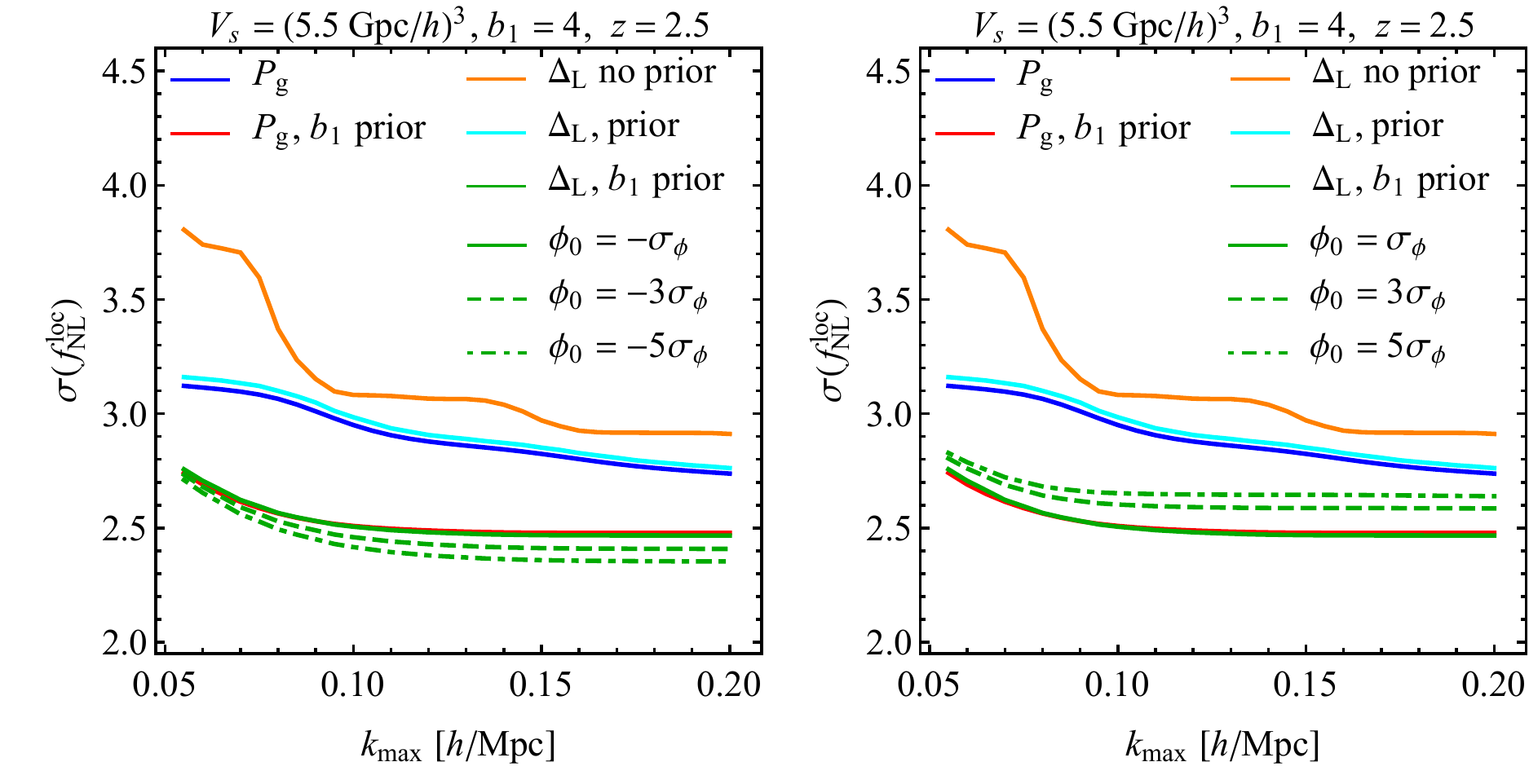}
    \caption{The error on $\fnl$ in the high-$z$ sample for different analysis choices. The blue line shows the standard case when no super-sample modes are considered. The red line present the effect of a prior on $b_1$. The impact of marginalizing over super-sample modes with or without a prior is shown by the orange and cyan lines. The green lines correspond to difference fiducial values of the large-scale gravitational potential $\phi_0$.}
    \label{fig:FisherQSO}
\end{figure}

The error on $\fnl$ as a function of $k_{\rm max}$, after marginalizing over the bias parameters and the amplitude of the long modes, is presented in \fig{fig:FisherQSO}. The standard case is shown in blue, and as widely known it exhibits a very weak dependence on the smallest scale included in the analysis \cite{Ferraro:2015}. If we include a prior on $b_1$, yielding a 40\% better measurement of linear bias compared to the error at $k_{\rm max} = 0.2\,\kMpc$, the constraint on $\fnl$ also improves by approximately 10\% at $k_{\rm max} = 0.2\,\kMpc$ (shown in red line).  Such a prior could arise from cross-correlation with other probes, \eg the CMB, other LSS tracers, or from the analysis of the bispectrum. 
The impact of marginalizing over the values of the long modes without assuming any prior on their values is shown as the orange line. For the orange line, the constraint is almost independent of the fiducial value $\Delta_0,\,\Delta_2$ and the value of $\phi_0$.  Therefore, we only plot the 1-$\sigma$ case for their fiducial value. The error on PNG is larger than the standard analysis, especially for low values of $k_{\rm max}\lesssim 0.1\,\kMpc$. For $k_{\rm max}\simeq 0.2\,\kMpc$ the degradation in the errorbar is less than a 10\%. Including a 3-$\sigma$ prior on the amplitude of the long modes results in the cyan line, which overlaps almost perfectly with the blue line. This is good news as we expect to be able to put strong theoretical priors on the value of the long wavelength modes. 

The most interesting case is when we put a prior on both $b_1$ and the amplitude of the long modes. The constraints on $\fnl$ with both priors are shown with green lines. In this scenario $\sigma(\fnl)$ depends on the fiducial value of $\phi_0$. This is easy to understand by noting that $\partial \hat{P}_g /\partial \fnl$ contains terms proportional to the value of the long modes. In particular since $R_\phi(k,z)$ is negative (see \fig{fig:zspace}) positive values of $\phi_0$ reduce the response of the galaxy power spectrum to PNG, while negative values enhances it. This was not manifest for the the cyan and orange line, when the marginalized constraint was dominated by the degeneracy between $b_1$ and $\fnl$. 

In the left panel the three green curves show the value of $\sigma({\fnl})$ for negative fiducial values of $\phi_0$. Continuous, dashed and dot-dashed lines correspond to $\phi_0 = -\{1,\,3,\,5\}$-$\sigma_\phi$. For large, therefore, more unlikely values of $\phi_0$ the constraint on $\fnl$ can be up to 10\% better than the case without long modes, shown in red. As expected lower, but more likely, values of $\phi_0$ show diminishing returns. For positive values of $\phi_0$ there are no significant improvements over the standard case by adding a prior on $b_1$, as shown in the right panel. This could potentially be a problem for multi-tracer analyses, as the benefits of a better measurement of linear bias could be hampered by the presence of super-sample modes.
\begin{figure}
    \centering
     \includegraphics[width=1\textwidth]{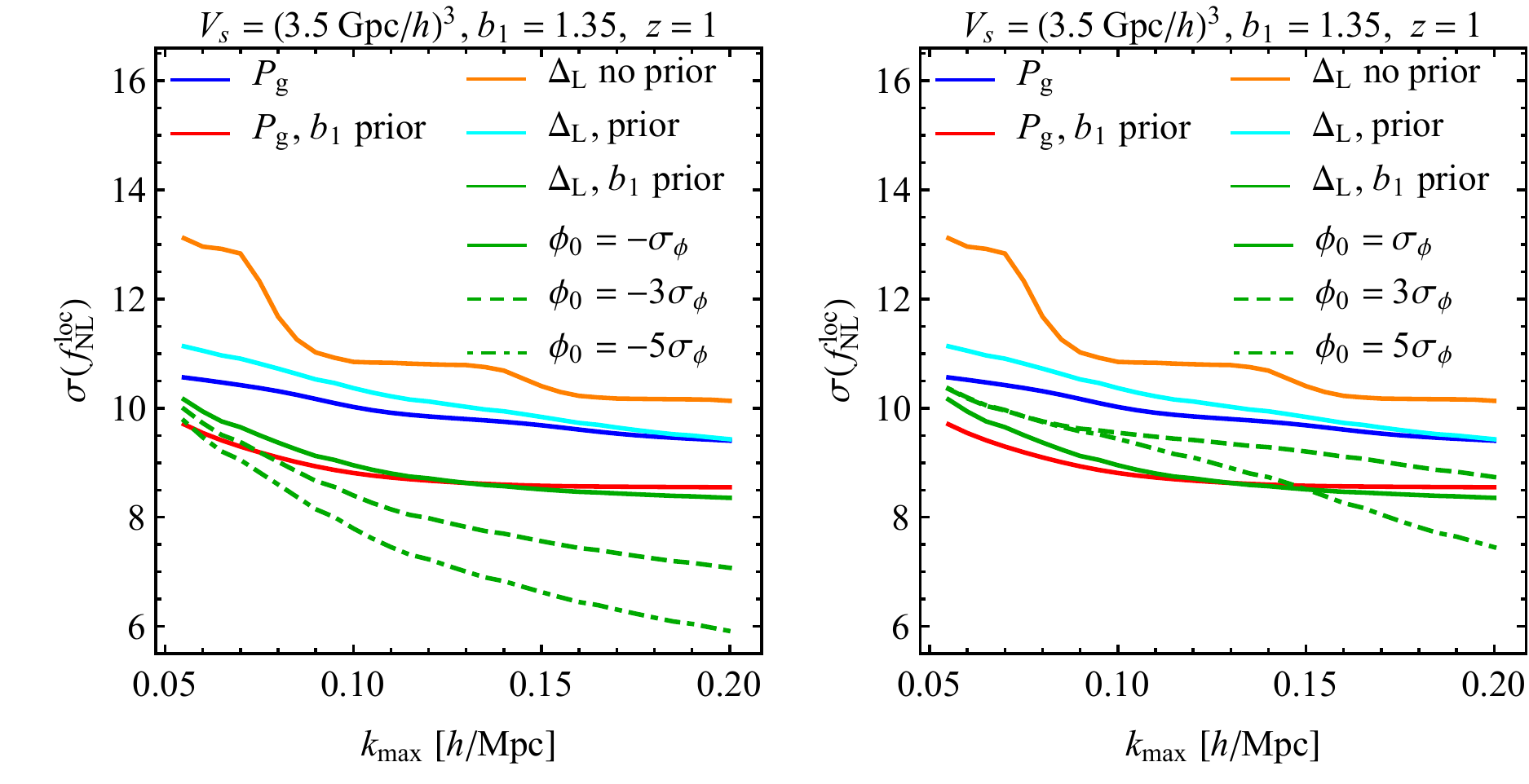}
    \caption{Same as \fig{fig:FisherQSO} but for the low-$z$ sample.}
    \label{fig:FisherELG}
\end{figure}

\subsection{The low-$z$, low bias sample}
The low-$z$ sample covers less volume and has a lower value of $b_1$, $b_\phi$ and $b_{\phi\delta}$ compared to the high-$z$ sample . We thus expect it to be less sensitive to local PNG. On the other hand, the lower value of the shot-noise compared to the high-$z$ sample means super-sample modes could contribute more to the total signal-to-noise of $\fnl$.
\fig{fig:FisherELG} shows the results using the same color coding of \fig{fig:FisherQSO}. The standard case is shown in blue, and we find the forecasted error on $\fnl$ is $\simeq$ 3 times worse than than in the high-$z$ sample. As before if we do not include any prior on the amplitude of the long modes, the constraint slightly degrades (the orange line), but even a very generous prior on $\Delta_L$ results in the cyan line which basically matches the blue one. It is important to notice that although the error on $\fnl$ is comparable between orange and the blue line, the linear bias $b_1$ is measured three times worse when super-sample modes are present. This is a because long wavelength modes mainly change the amplitude of the galaxy power spectrum; therefore they are very degenerate with the linear bias.

It is thus interesting to see how imposing a prior on linear bias, making the constraint on $b_1$ is similar with and without the long modes, could affect the constraint on PNG. We then include a prior on $b_1$ such that $\sigma(b_1)$ at $k_{\rm max} =0.2 \,\kMpc$ becomes 40\% better than the standard case (the blue line), yielding the same measurement of linear bias irrespective of the presence of the super sample fluctuations. In the absence of the long modes, the constraint on $\fnl$, shown with a red line, improves by roughly 10\% at the highest $k_{\rm max}$. When we include the super-sample modes, the improvement due to the prior on $b_1$ varies between (10 - 30)\% for negative values of $\phi_0$, with the rarest negative $5\sigma$ fluctuation yielding more than 30\% better constraints on $\fnl$. This was possible due the low shot noise level of the low-$z$ sample. The picture is somewhat reversed for positive values of $\phi_0$, where we do not find significant improvement over the standard case in the presence of super-sample signal.

\section{Conclusions}
\label{sec:conclude}
In this paper we investigated the effect of super-survey modes on the galaxy power spectrum in the presence of primordial non-Gaussianity.
We extended previous work in real-space and computed for the first time the response of the multipoles of the redshift-space power spectrum to the super-sample gravitational potential. We also clarified the role of the stochastic terms in the bias expansion when computing such responses. PNG generate new couplings between the small-scale gravitational potential and the isotropic and tidal part of the super-sample modes, as well as correlations between the large-scale gravitational potential and the small-scale fluctuations. The former are a specific feature of biased tracer and would be zero for dark matter, while the latter is generic outcome for all fields.

We then forecasted the effect of super-sample modes on the determination of $\fnl$, including their contribution to both the signal and the noise part of the covariance of the redshift-space multipoles of the galaxy power spectrum. Focusing on two hypothetical galaxy samples, one at low-$z$ and one at high-$z$, we find that the addition of the long modes as new free parameters, without any prior, degrades $\sigma({\fnl})$ by roughly 10\% at high-$k$ compared to the standard Fisher forecast that neglects the super-sample effects. A mild prior on the long wavelength fluctuations is able to recover the constraint in the standard scenario. We find that although the error on $\fnl$ is similar with and without long modes, the linear bias is measured three times worse in the latter case. We therefore studied a scenario where a prior on $b_1$ makes the linear bias measured to a similar precision independently of the presence of super-sample modes. In this case we find that negative values of the long wavelength gravitational potential $\phi_0$ yields smaller $\sigma(\fnl)$ compared to the  $\phi_0\geq0$ cosmology. For samples with a high enough number density, like the DESI ELG sample, the improvement can be up to 30\% for large negative values of $\phi_0$. However, it is worth reminding that such values are the most unlikely to be realized. 

A number of simplifying assumptions have been made in this work, and they would have to be addressed before application of our formalism to real data is possible. For instance a varying LOS in curved sky analysis, more complicated window functions, or the so called radial integral constraint \cite{deMattia:2019vdg}. We leave those to future work, but anticipate that they would not lead to qualitative changes in the results presented in this paper.

 \acknowledgments{EC would like to thank Yin Li,
Jay Wadekar and Roman Scoccimarro for useful discussions. A.M.D. is supported by the SNSF project, {\it The Non-Gaussian Universe and Cosmological Symmetries}, project number:200020-178787.}
 
\appendix
\section{Tree-level galaxy bispectrum}\label{app:bis_tree}
\subsection{Real-space}
Assuming statistical isotropy and using the bias expansion in Eq. \eqref{eq:bias}, the galaxy bispectrum at tree-level can be written as 
\begin{equation}
B_g(k_1,k_2,k_3,z) = B_g^G(k_1,k_2,k_3,z) + B_g^{\rm NG}(k_1,k_2,k_3,z),
\end{equation}
where the first contribution is induced by gravitational evolution and is given by 
\begin{equation}
B_g^G(k_1,k_2,k_3) = 2 b_1^2 \left[b_1 F_2({\bf k}_1,{\bf k}_2) + b_2 + b_{s^2}s^2({\bf k}_1,{\bf k}_2)\right]P(k_1,z)P(k_2,z) + 2\ {\rm perms},
\end{equation}
while the second contribution is due to primordial non-Gaussianity and is given by \cite{Desjacques:2016}
\begin{align}
 B^{\rm NG}_g(k_1,k_2,k_3,z) &=  b_1^3 B(k_1,k_2,k_3,z) + \fnl \Bigg\{b_1^2  \ b_\phi \left[\frac{k_1}{k_2} {\mathcal M}^{-1}(k_1,z) + \frac{k_2}{k_1} {\mathcal M}^{-1}(k_2,z)\right] \mu_{12} \ \Bigg. \nonumber \\
&+ 2 b_1 b_\phi \left[{\mathcal M}^{-1}(k_1,z) + {\mathcal M}^{-1}(k_2,z)\right]  \left[ b_1 F_2({\bf k}_1,{\bf k}_2) + b_2 + b_{s^2}s^2({\bf k}_1,{\bf k}_2)\right]  \nonumber \\
&+ \Bigg. b_1^2 \ b_{\phi\delta} \left[{\mathcal M}^{-1}(k_1,z) + {\mathcal M}^{-1}(k_2,z)\right] \Bigg\}P(k_1,z)P(k_2,z) + 2 \ {\rm perms.}
\end{align}
Here, $\mu_{12} = \hat {\bf k}_1.\hat {\bf k}_2$ is the angle between the two wavevectors $\bf k_1$ and $\bf k_2$, and $F_2$ is the second-order kernel in standard perturbation theory,
\begin{equation}
 F_2({\bf k}_1,{\bf k}_2) = \frac{5}{7} + \frac{1}{2} \left(\frac{k_1}{k_2} + \frac{k_2}{k_1}\right) \mu_{12}+ \frac{2}{7}\ \mu_{12}^2\, ,
\end{equation}
and $B$ is the linear matter bispectrum that is sourced by non-zero primordial bispectrum
\begin{equation}
B(k_1,k_2,k_3,z) = {\mathcal M}(k_1,z){\mathcal M}(k_2,z) {\mathcal M}(k_3,z)  B^{\rm loc}_\phi(k_1,k_2,k_3)\,.
\end{equation}
with
\begin{equation} 
B_\phi^{\rm loc}(k_1,k_2,k_3) = 2 \fnl \left[P_\phi(k_1)P_\phi(k_2) + 2\ {\rm perms} \right].  
\end{equation}

\subsection{Redshift-space}
In redshift-space, the bispectrum depends on 5 variables, which we can choose to be three sides of the triangles and two angles to define the position of the triangles with respect to the line of sight. At tree-level in perturbation theory, and including primordial non-Gaussianity, the bispectrum is given by 

\begin{align}\label{eq:Bgrav}
	\hskip -6.5pt B_g (\k_1,\k_2,\k_3,z) &= Z_1(\k_1)Z_1(\k_2)Z_1(\k_2) B(k_1,k_2,k_3) \notag \\
	&+ \left\{2 Z_1(\k_1) Z_1(\k_2) Z_2(\k_1,\k_2) P(k_1,z)P(k_2,z) + 2 \ \text{perms}\right\}\, ,
\end{align}
where
\begin{align}
\label{eq:Zfunction}
	Z_1(\k_1) &= b_1+f \mu_1^2 + \fnl b_\phi {\mathcal M}^{-1}(k_1,z)\, ,\nonumber \\
	Z_2(\k_1,\k_2) &= \frac{b_2}{2} + b_1 F_2(\k_1,\k_2) + f\mu_3^2\, G_2(\k_1,\k_2) \nonumber \\
	&- \frac{f \mu_3 k_3}{2}\left[ \frac{\mu_1}{k_1} Z_1(\k_2) +\frac{\mu_2}{k_2}Z_1(\k_1)\right] + b_{s^2}s^2(\k_1,\k_2) \notag \\
	&+ b_\phi \fnl \left[\frac{k_1}{k_2}{\mathcal M}^{-1}(k_1,z) + \frac{k_2}{k_1}{\mathcal M}^{-1}(k_2,z) \right] \mu_{12}\notag \\
	&+ b_{\phi \delta} \fnl \left[{\mathcal M}^{-1}(k_1,z) + {\mathcal M}^{-1}(k_2,z)\right] \, ,
\end{align}
with $\mu_i = \hat \k_i.\hat {\bf n}$ being the angles between a given wavevector and line-of-sight direction, and $G_2$ is the second-order kernel of matter velocity contrast
\begin{align}
	G_2(\k_1,\k_2) &\equiv \frac{3}{7} + \frac{1}{2}\left(\frac{k_1}{k_2} + \frac{k_2}{k_1}\right) \mu_{12} + \frac{4}{7} \mu_{12}^2\, .
\end{align}
Note that in deriving the response functions, we only keep the terms linear in $\fnl$. Compared to the similar expressions in Ref. \cite{Karagiannis}, we have an additional contribution to $Z_2$ kernel, which is due to the transformation of the gravitational potential $\phi$ from Lagrangian to Eulerian coordinates.

\bibliographystyle{JHEP}
\bibliography{}

\providecommand{\href}[2]{#2}\begingroup\raggedright\begin{thebibliography}{10}

\bibitem{Kaiser:1987qv}
N.~Kaiser, \emph{{Clustering in real space and in redshift space}}, {\emph{Mon.
  Not. Roy. Astron. Soc.} {\bf 227} (1987) 1--27}.

\bibitem{Hamilton:2005dx}
A.~J. Hamilton, C.~D. Rimes and R.~Scoccimarro, \emph{{On measuring the
  covariance matrix of the nonlinear power spectrum from simulations}},
  \href{http://dx.doi.org/10.1111/j.1365-2966.2006.10709.x}{\emph{Mon. Not.
  Roy. Astron. Soc.} {\bf 371} (2006) 1188--1204},
  [\href{https://arxiv.org/abs/astro-ph/0511416}{{\tt astro-ph/0511416}}].

\bibitem{Sirko:2005uz}
E.~Sirko, \emph{{Initial conditions to cosmological N-body simulations, or how
  to run an ensemble of simulations}},
  \href{http://dx.doi.org/10.1086/497090}{\emph{Astrophys. J.} {\bf 634} (2005)
  728--743}, [\href{https://arxiv.org/abs/astro-ph/0503106}{{\tt
  astro-ph/0503106}}].

\bibitem{Hu:2002we}
W.~Hu and A.~V. Kravtsov, \emph{{Sample variance considerations for cluster
  surveys}}, \href{http://dx.doi.org/10.1086/345846}{\emph{Astrophys. J.} {\bf
  584} (2003) 702--715}, [\href{https://arxiv.org/abs/astro-ph/0203169}{{\tt
  astro-ph/0203169}}].

\bibitem{dePutter:2011ah}
R.~de~Putter, C.~Wagner, O.~Mena, L.~Verde and W.~Percival, \emph{{Thinking
  Outside the Box: Effects of Modes Larger than the Survey on Matter Power
  Spectrum Covariance}},
  \href{http://dx.doi.org/10.1088/1475-7516/2012/04/019}{\emph{JCAP} {\bf 04}
  (2012) 019}, [\href{https://arxiv.org/abs/1111.6596}{{\tt 1111.6596}}].

\bibitem{Baldauf:2011bh}
T.~Baldauf, U.~Seljak, L.~Senatore and M.~Zaldarriaga, \emph{{Galaxy Bias and
  non-Linear Structure Formation in General Relativity}},
  \href{http://dx.doi.org/10.1088/1475-7516/2011/10/031}{\emph{JCAP} {\bf 10}
  (2011) 031}, [\href{https://arxiv.org/abs/1106.5507}{{\tt 1106.5507}}].

\bibitem{Takada:2013wfa}
M.~Takada and W.~Hu, \emph{{Power Spectrum Super-Sample Covariance}},
  \href{http://dx.doi.org/10.1103/PhysRevD.87.123504}{\emph{Phys. Rev. D} {\bf
  87} (2013) 123504}, [\href{https://arxiv.org/abs/1302.6994}{{\tt
  1302.6994}}].

\bibitem{Wagner:2015gva}
C.~Wagner, F.~Schmidt, C.-T. Chiang and E.~Komatsu, \emph{{The angle-averaged
  squeezed limit of nonlinear matter N-point functions}},
  \href{http://dx.doi.org/10.1088/1475-7516/2015/08/042}{\emph{JCAP} {\bf 08}
  (2015) 042}, [\href{https://arxiv.org/abs/1503.03487}{{\tt 1503.03487}}].

\bibitem{Li:2014jra}
Y.~Li, W.~Hu and M.~Takada, \emph{{Super-Sample Signal}},
  \href{http://dx.doi.org/10.1103/PhysRevD.90.103530}{\emph{Phys. Rev. D} {\bf
  90} (2014) 103530}, [\href{https://arxiv.org/abs/1408.1081}{{\tt
  1408.1081}}].

\bibitem{Martino:2009dm}
M.~C. Martino and R.~K. Sheth, \emph{{On the equivalence between the effective
  cosmology and excursion set treatments of environment}},
  \href{http://dx.doi.org/10.1111/j.1365-2966.2009.14467.x}{\emph{Mon. Not.
  Roy. Astron. Soc.} {\bf 394} (2009) 2109},
  [\href{https://arxiv.org/abs/0901.0757}{{\tt 0901.0757}}].

\bibitem{Seljak:2012tp}
U.~Seljak, \emph{{Bias, redshift space distortions and primordial
  nongaussianity of nonlinear transformations: application to Lyman alpha
  forest}}, \href{http://dx.doi.org/10.1088/1475-7516/2012/03/004}{\emph{JCAP}
  {\bf 03} (2012) 004}, [\href{https://arxiv.org/abs/1201.0594}{{\tt
  1201.0594}}].

\bibitem{Li:2014sga}
Y.~Li, W.~Hu and M.~Takada, \emph{{Super-Sample Covariance in Simulations}},
  \href{http://dx.doi.org/10.1103/PhysRevD.89.083519}{\emph{Phys. Rev. D} {\bf
  89} (2014) 083519}, [\href{https://arxiv.org/abs/1401.0385}{{\tt
  1401.0385}}].

\bibitem{Chiang:2014oga}
C.-T. Chiang, C.~Wagner, F.~Schmidt and E.~Komatsu, \emph{{Position-dependent
  power spectrum of the large-scale structure: a novel method to measure the
  squeezed-limit bispectrum}},
  \href{http://dx.doi.org/10.1088/1475-7516/2014/05/048}{\emph{JCAP} {\bf 05}
  (2014) 048}, [\href{https://arxiv.org/abs/1403.3411}{{\tt 1403.3411}}].

\bibitem{Feix:2013hha}
M.~Feix and A.~Nusser, \emph{{Beyond boundaries of redshift surveys: assessing
  mass fluctuations on "super-survey" scales}},
  \href{http://dx.doi.org/10.1088/1475-7516/2013/12/027}{\emph{JCAP} {\bf 12}
  (2013) 027}, [\href{https://arxiv.org/abs/1306.4719}{{\tt 1306.4719}}].

\bibitem{Barreira:2017sqa}
A.~Barreira and F.~Schmidt, \emph{{Responses in Large-Scale Structure}},
  \href{http://dx.doi.org/10.1088/1475-7516/2017/06/053}{\emph{JCAP} {\bf 06}
  (2017) 053}, [\href{https://arxiv.org/abs/1703.09212}{{\tt 1703.09212}}].

\bibitem{Schaan:2014cpa}
E.~Schaan, M.~Takada and D.~N. Spergel, \emph{{Joint likelihood function of
  cluster counts and $n$-point correlation functions: Improving their power
  through including halo sample variance}},
  \href{http://dx.doi.org/10.1103/PhysRevD.90.123523}{\emph{Phys. Rev. D} {\bf
  90} (2014) 123523}, [\href{https://arxiv.org/abs/1406.3330}{{\tt
  1406.3330}}].

\bibitem{Bertolini:2016hxg}
D.~Bertolini and M.~P. Solon, \emph{{Principal Shapes and Squeezed Limits in
  the Effective Field Theory of Large Scale Structure}},
  \href{http://dx.doi.org/10.1088/1475-7516/2016/11/030}{\emph{JCAP} {\bf 11}
  (2016) 030}, [\href{https://arxiv.org/abs/1608.01310}{{\tt 1608.01310}}].

\bibitem{McDonald:2001fe}
P.~McDonald, \emph{{Toward a measurement of the cosmological geometry at Z 2:
  predicting lyman-alpha forest correlation in three dimensions, and the
  potential of future data sets}},
  \href{http://dx.doi.org/10.1086/345945}{\emph{Astrophys. J.} {\bf 585} (2003)
  34--51}, [\href{https://arxiv.org/abs/astro-ph/0108064}{{\tt
  astro-ph/0108064}}].

\bibitem{Li:2015jsz}
Y.~Li, W.~Hu and M.~Takada, \emph{{Separate Universe Consistency Relation and
  Calibration of Halo Bias}},
  \href{http://dx.doi.org/10.1103/PhysRevD.93.063507}{\emph{Phys. Rev. D} {\bf
  93} (2016) 063507}, [\href{https://arxiv.org/abs/1511.01454}{{\tt
  1511.01454}}].

\bibitem{Baldauf:2015vio}
T.~Baldauf, U.~s. Seljak, L.~Senatore and M.~Zaldarriaga, \emph{{Linear
  response to long wavelength fluctuations using curvature simulations}},
  \href{http://dx.doi.org/10.1088/1475-7516/2016/09/007}{\emph{JCAP} {\bf 09}
  (2016) 007}, [\href{https://arxiv.org/abs/1511.01465}{{\tt 1511.01465}}].

\bibitem{Lazeyras:2015lgp}
T.~Lazeyras, C.~Wagner, T.~Baldauf and F.~Schmidt, \emph{{Precision measurement
  of the local bias of dark matter halos}},
  \href{http://dx.doi.org/10.1088/1475-7516/2016/02/018}{\emph{JCAP} {\bf 02}
  (2016) 018}, [\href{https://arxiv.org/abs/1511.01096}{{\tt 1511.01096}}].

\bibitem{Modi2016}
C.~Modi, E.~Castorina and U.~Seljak, \emph{{Halo bias in Lagrangian Space:
  Estimators and theoretical predictions}},
  \href{http://dx.doi.org/10.1093/mnras/stx2148}{\emph{Mon. Not. Roy. Astron.
  Soc.} {\bf 472} (2017) 3959--3970},
  [\href{https://arxiv.org/abs/1612.01621}{{\tt 1612.01621}}].

\bibitem{Akitsu:2016leq}
K.~Akitsu, M.~Takada and Y.~Li, \emph{{Large-scale tidal effect on
  redshift-space power spectrum in a finite-volume survey}},
  \href{http://dx.doi.org/10.1103/PhysRevD.95.083522}{\emph{Phys. Rev. D} {\bf
  95} (2017) 083522}, [\href{https://arxiv.org/abs/1611.04723}{{\tt
  1611.04723}}].

\bibitem{Akitsu:2017syq}
K.~Akitsu and M.~Takada, \emph{{Impact of large-scale tides on cosmological
  distortions via redshift-space power spectrum}},
  \href{http://dx.doi.org/10.1103/PhysRevD.97.063527}{\emph{Phys. Rev. D} {\bf
  97} (2018) 063527}, [\href{https://arxiv.org/abs/1711.00012}{{\tt
  1711.00012}}].

\bibitem{Chiang:2018mau}
C.-T. Chiang and A.~z. Slosar, \emph{{Power spectrum in the presence of
  large-scale overdensity and tidal fields: breaking azimuthal symmetry}},
  \href{http://dx.doi.org/10.1088/1475-7516/2018/07/049}{\emph{JCAP} {\bf 07}
  (2018) 049}, [\href{https://arxiv.org/abs/1804.02753}{{\tt 1804.02753}}].

\bibitem{Li:2017qgh}
Y.~Li, M.~Schmittfull and U.~s. Seljak, \emph{{Galaxy power-spectrum responses
  and redshift-space super-sample effect}},
  \href{http://dx.doi.org/10.1088/1475-7516/2018/02/022}{\emph{JCAP} {\bf 02}
  (2018) 022}, [\href{https://arxiv.org/abs/1711.00018}{{\tt 1711.00018}}].

\bibitem{Stucker:2020fhk}
J.~Stücker, A.~Schmidt, S.~D. White, F.~Schmidt and O.~Hahn, \emph{{Measuring
  the Tidal Response of Structure Formation: Anisotropic Separate Universe
  Simulations using TreePM}},  \href{https://arxiv.org/abs/2003.06427}{{\tt
  2003.06427}}.

\bibitem{Schmidt:2018hbj}
A.~S. Schmidt, S.~D. White, F.~Schmidt and J.~Stücker, \emph{{Cosmological
  N-Body Simulations with a Large-Scale Tidal Field}},
  \href{http://dx.doi.org/10.1093/mnras/sty1430}{\emph{Mon. Not. Roy. Astron.
  Soc.} {\bf 479} (2018) 162--170},
  [\href{https://arxiv.org/abs/1803.03274}{{\tt 1803.03274}}].

\bibitem{Masaki:2020drx}
S.~Masaki, T.~Nishimichi and M.~Takada, \emph{{Anisotropic separate universe
  simulations}},  \href{https://arxiv.org/abs/2003.10052}{{\tt 2003.10052}}.

\bibitem{Mohammed:2014lja}
I.~Mohammed and U.~Seljak, \emph{{Analytic model for the matter power spectrum,
  its covariance matrix, and baryonic effects}},
  \href{http://dx.doi.org/10.1093/mnras/stu1972}{\emph{Mon. Not. Roy. Astron.
  Soc.} {\bf 445} (2014) 3382--3400},
  [\href{https://arxiv.org/abs/1407.0060}{{\tt 1407.0060}}].

\bibitem{Chan:2017fiv}
K.~C. Chan, A.~Moradinezhad~Dizgah and J.~Noreña, \emph{{Bispectrum
  Supersample Covariance}},
  \href{http://dx.doi.org/10.1103/PhysRevD.97.043532}{\emph{Phys. Rev. D} {\bf
  97} (2018) 043532}, [\href{https://arxiv.org/abs/1709.02473}{{\tt
  1709.02473}}].

\bibitem{Wadekar2019}
D.~Wadekar and R.~Scoccimarro, \emph{{The Galaxy Power Spectrum Multipoles
  Covariance in Perturbation Theory}},
  \href{https://arxiv.org/abs/1910.02914}{{\tt 1910.02914}}.

\bibitem{Alvarez:2014}
M.~{Alvarez}, T.~{Baldauf}, J.~R. {Bond}, N.~{Dalal}, R.~{de Putter},
  O.~{Dor{\'e}} et~al., \emph{{Testing Inflation with Large Scale Structure:
  Connecting Hopes with Reality}}, {\emph{ArXiv e-prints} (Dec., 2014) },
  [\href{https://arxiv.org/abs/1412.4671}{{\tt 1412.4671}}].

\bibitem{Abazajian:2016}
K.~N. {Abazajian}, P.~{Adshead}, Z.~{Ahmed}, S.~W. {Allen}, D.~{Alonso}, K.~S.
  {Arnold} et~al., \emph{{CMB-S4 Science Book, First Edition}}, {\emph{ArXiv
  e-prints} (Oct., 2016) }, [\href{https://arxiv.org/abs/1610.02743}{{\tt
  1610.02743}}].

\bibitem{Munchmeyer:2018eey}
M.~Münchmeyer, M.~S. Madhavacheril, S.~Ferraro, M.~C. Johnson and K.~M. Smith,
  \emph{{Constraining local non-Gaussianities with kinetic Sunyaev-Zel'dovich
  tomography}},
  \href{http://dx.doi.org/10.1103/PhysRevD.100.083508}{\emph{Phys. Rev. D} {\bf
  100} (2019) 083508}, [\href{https://arxiv.org/abs/1810.13424}{{\tt
  1810.13424}}].

\bibitem{Tellarini:2016sgp}
M.~Tellarini, A.~J. Ross, G.~Tasinato and D.~Wands, \emph{{Galaxy bispectrum,
  primordial non-Gaussianity and redshift space distortions}},
  \href{http://dx.doi.org/10.1088/1475-7516/2016/06/014}{\emph{JCAP} {\bf 06}
  (2016) 014}, [\href{https://arxiv.org/abs/1603.06814}{{\tt 1603.06814}}].

\bibitem{Karagiannis}
D.~{Karagiannis}, A.~{Lazanu}, M.~{Liguori}, A.~{Raccanelli}, N.~{Bartolo} and
  L.~{Verde}, \emph{{Constraining primordial non-Gaussianity with bispectrum
  and power spectrum from upcoming optical and radio surveys}},
  \href{http://dx.doi.org/10.1093/mnras/sty1029}{\emph{\mnras} {\bf 478} (July,
  2018) 1341--1376}, [\href{https://arxiv.org/abs/1801.09280}{{\tt
  1801.09280}}].

\bibitem{MoradinezhadDizgah:2018pfo}
A.~Moradinezhad~Dizgah, G.~Franciolini, A.~Kehagias and A.~Riotto,
  \emph{{Constraints on long-lived, higher-spin particles from galaxy
  bispectrum}}, \href{http://dx.doi.org/10.1103/PhysRevD.98.063520}{\emph{Phys.
  Rev. D} {\bf 98} (2018) 063520},
  [\href{https://arxiv.org/abs/1805.10247}{{\tt 1805.10247}}].

\bibitem{MoradinezhadDizgah:2018ssw}
A.~Moradinezhad~Dizgah, H.~Lee, J.~B. Muñoz and C.~Dvorkin, \emph{{Galaxy
  Bispectrum from Massive Spinning Particles}},
  \href{http://dx.doi.org/10.1088/1475-7516/2018/05/013}{\emph{JCAP} {\bf 05}
  (2018) 013}, [\href{https://arxiv.org/abs/1801.07265}{{\tt 1801.07265}}].

\bibitem{Gualdi:2020ymf}
D.~Gualdi and L.~Verde, \emph{{Galaxy redshift-space bispectrum: the Importance
  of Being Anisotropic}},  \href{https://arxiv.org/abs/2003.12075}{{\tt
  2003.12075}}.

\bibitem{Camera:2013kpa}
S.~Camera, M.~G. Santos, P.~G. Ferreira and L.~Ferramacho, \emph{{Cosmology on
  Ultra-Large Scales with HI Intensity Mapping: Limits on Primordial
  non-Gaussianity}},
  \href{http://dx.doi.org/10.1103/PhysRevLett.111.171302}{\emph{Phys. Rev.
  Lett.} {\bf 111} (2013) 171302}, [\href{https://arxiv.org/abs/1305.6928}{{\tt
  1305.6928}}].

\bibitem{MoradinezhadDizgah:2018lac}
A.~Moradinezhad~Dizgah and G.~K. Keating, \emph{{Line intensity mapping with
  [CII] and CO(1-0) as probes of primordial non-Gaussianity}},
  \href{http://dx.doi.org/10.3847/1538-4357/aafd36}{\emph{Astrophys. J.} {\bf
  872} (2019) 126}, [\href{https://arxiv.org/abs/1810.02850}{{\tt
  1810.02850}}].

\bibitem{Karagiannis:2019jjx}
D.~Karagiannis, A.~z. Slosar and M.~Liguori, \emph{{Forecasts on Primordial
  non-Gaussianity from 21 cm Intensity Mapping experiments}},
  \href{https://arxiv.org/abs/1911.03964}{{\tt 1911.03964}}.

\bibitem{PlanckFNL}
{\scshape Planck} collaboration, Y.~Akrami et~al., \emph{{Planck 2018 results.
  IX. Constraints on primordial non-Gaussianity}},
  \href{https://arxiv.org/abs/1905.05697}{{\tt 1905.05697}}.

\bibitem{Maldacena:2003}
J.~{Maldacena}, \emph{{Non-gaussian features of primordial fluctuations in
  single field inflationary models}},
  \href{http://dx.doi.org/10.1088/1126-6708/2003/05/013}{\emph{Journal of High
  Energy Physics} {\bf 5} (May, 2003) 013},
  [\href{https://arxiv.org/abs/astro-ph/0210603}{{\tt astro-ph/0210603}}].

\bibitem{Creminelli:2004}
P.~{Creminelli} and M.~{Zaldarriaga}, \emph{{A single-field consistency
  relation for the three-point function}},
  \href{http://dx.doi.org/10.1088/1475-7516/2004/10/006}{\emph{\jcap} {\bf 10}
  (Oct., 2004) 006}, [\href{https://arxiv.org/abs/astro-ph/0407059}{{\tt
  astro-ph/0407059}}].

\bibitem{JeongKomatsu}
D.~{Jeong} and E.~{Komatsu}, \emph{{Primordial Non-Gaussianity, Scale-dependent
  Bias, and the Bispectrum of Galaxies}},
  \href{http://dx.doi.org/10.1088/0004-637X/703/2/1230}{\emph{\apj} {\bf 703}
  (Oct., 2009) 1230--1248}, [\href{https://arxiv.org/abs/0904.0497}{{\tt
  0904.0497}}].

\bibitem{Sefusatti2009}
E.~{Sefusatti}, \emph{{One-loop perturbative corrections to the matter and
  galaxy bispectrum with non-Gaussian initial conditions}},
  \href{http://dx.doi.org/10.1103/PhysRevD.80.123002}{\emph{\prd} {\bf 80}
  (Dec., 2009) 123002}, [\href{https://arxiv.org/abs/0905.0717}{{\tt
  0905.0717}}].

\bibitem{Baldauf11}
T.~{Baldauf}, U.~{Seljak} and L.~{Senatore}, \emph{{Primordial non-Gaussianity
  in the bispectrum of the halo density field}},
  \href{http://dx.doi.org/10.1088/1475-7516/2011/04/006}{\emph{\jcap} {\bf
  2011} (Apr., 2011) 006}, [\href{https://arxiv.org/abs/1011.1513}{{\tt
  1011.1513}}].

\bibitem{Sefusatti:2011gt}
E.~Sefusatti, M.~Crocce and V.~Desjacques, \emph{{The Halo Bispectrum in N-body
  Simulations with non-Gaussian Initial Conditions}},
  \href{http://dx.doi.org/10.1111/j.1365-2966.2012.21271.x}{\emph{Mon. Not.
  Roy. Astron. Soc.} {\bf 425} (2012) 2903},
  [\href{https://arxiv.org/abs/1111.6966}{{\tt 1111.6966}}].

\bibitem{Dalal:2008}
N.~{Dalal}, O.~{Dor{\'e}}, D.~{Huterer} and A.~{Shirokov}, \emph{{Imprints of
  primordial non-Gaussianities on large-scale structure: Scale-dependent bias
  and abundance of virialized objects}},
  \href{http://dx.doi.org/10.1103/PhysRevD.77.123514}{\emph{\prd} {\bf 77}
  (June, 2008) 123514}, [\href{https://arxiv.org/abs/0710.4560}{{\tt
  0710.4560}}].

\bibitem{Slosar:2008}
A.~{Slosar}, C.~{Hirata}, U.~{Seljak}, S.~{Ho} and N.~{Padmanabhan},
  \emph{{Constraints on local primordial non-Gaussianity from large scale
  structure}},
  \href{http://dx.doi.org/10.1088/1475-7516/2008/08/031}{\emph{\jcap} {\bf 8}
  (Aug., 2008) 031}, [\href{https://arxiv.org/abs/0805.3580}{{\tt 0805.3580}}].

\bibitem{Matarrese:2008}
S.~{Matarrese} and L.~{Verde}, \emph{{The Effect of Primordial Non-Gaussianity
  on Halo Bias}}, \href{http://dx.doi.org/10.1086/587840}{\emph{\apjl} {\bf
  677} (Apr., 2008) L77}, [\href{https://arxiv.org/abs/0801.4826}{{\tt
  0801.4826}}].

\bibitem{Afshordi:2008ru}
N.~Afshordi and A.~J. Tolley, \emph{{Primordial non-gaussianity, statistics of
  collapsed objects, and the Integrated Sachs-Wolfe effect}},
  \href{http://dx.doi.org/10.1103/PhysRevD.78.123507}{\emph{Phys. Rev. D} {\bf
  78} (2008) 123507}, [\href{https://arxiv.org/abs/0806.1046}{{\tt
  0806.1046}}].

\bibitem{Leistedt:2014}
B.~{Leistedt}, H.~V. {Peiris} and N.~{Roth}, \emph{{Constraints on Primordial
  Non-Gaussianity from 800 000 Photometric Quasars}},
  \href{http://dx.doi.org/10.1103/PhysRevLett.113.221301}{\emph{Physical Review
  Letters} {\bf 113} (Nov., 2014) 221301},
  [\href{https://arxiv.org/abs/1405.4315}{{\tt 1405.4315}}].

\bibitem{Giannantonio:2014b}
T.~{Giannantonio}, A.~J. {Ross}, W.~J. {Percival}, R.~{Crittenden},
  D.~{Bacher}, M.~{Kilbinger} et~al., \emph{{Improved primordial
  non-Gaussianity constraints from measurements of galaxy clustering and the
  integrated Sachs-Wolfe effect}},
  \href{http://dx.doi.org/10.1103/PhysRevD.89.023511}{\emph{\prd} {\bf 89}
  (Jan., 2014) 023511}, [\href{https://arxiv.org/abs/1303.1349}{{\tt
  1303.1349}}].

\bibitem{Castorina2019}
E.~Castorina et~al., \emph{{Redshift-weighted constraints on primordial
  non-Gaussianity from the clustering of the eBOSS DR14 quasars in Fourier
  space}}, \href{http://dx.doi.org/10.1088/1475-7516/2019/09/010}{\emph{JCAP}
  {\bf 09} (2019) 010}, [\href{https://arxiv.org/abs/1904.08859}{{\tt
  1904.08859}}].

\bibitem{DESI:2016}
{DESI Collaboration}, A.~{Aghamousa}, J.~{Aguilar}, S.~{Ahlen}, S.~{Alam},
  L.~E. {Allen} et~al., \emph{{The DESI Experiment Part I: Science,Targeting,
  and Survey Design}}, {\emph{ArXiv e-prints} (Oct., 2016) },
  [\href{https://arxiv.org/abs/1611.00036}{{\tt 1611.00036}}].

\bibitem{Dore:2014}
O.~{Dor{\'e}}, J.~{Bock}, M.~{Ashby}, P.~{Capak}, A.~{Cooray}, R.~{de Putter}
  et~al., \emph{{Cosmology with the SPHEREX All-Sky Spectral Survey}},
  {\emph{ArXiv e-prints} (Dec., 2014) },
  [\href{https://arxiv.org/abs/1412.4872}{{\tt 1412.4872}}].

\bibitem{Seljak:2009}
U.~{Seljak}, \emph{{Extracting Primordial Non-Gaussianity without Cosmic
  Variance}},
  \href{http://dx.doi.org/10.1103/PhysRevLett.102.021302}{\emph{Physical Review
  Letters} {\bf 102} (Jan., 2009) 021302},
  [\href{https://arxiv.org/abs/0807.1770}{{\tt 0807.1770}}].

\bibitem{Castorina_zerobias}
E.~{Castorina}, Y.~{Feng}, U.~{Seljak} and F.~{Villaescusa-Navarro},
  \emph{{Primordial non-Gaussianities and zero bias tracers of the Large Scale
  Structure}}, {\emph{ArXiv e-prints} (Mar., 2018) },
  [\href{https://arxiv.org/abs/1803.11539}{{\tt 1803.11539}}].

\bibitem{Ginzburg:2019xsj}
D.~Ginzburg and V.~Desjacques, \emph{{Shot noise in multi-tracer constraints on
  $f_\text{NL}$ and relativistic projections: Power Spectrum}},
  \href{https://arxiv.org/abs/1911.11701}{{\tt 1911.11701}}.

\bibitem{Chiang:2017vuk}
C.-T. Chiang, W.~Hu, Y.~Li and M.~Loverde, \emph{{Scale-dependent bias and
  bispectrum in neutrino separate universe simulations}},
  \href{http://dx.doi.org/10.1103/PhysRevD.97.123526}{\emph{Phys. Rev. D} {\bf
  97} (2018) 123526}, [\href{https://arxiv.org/abs/1710.01310}{{\tt
  1710.01310}}].

\bibitem{dePutter:2018}
R.~de~Putter, \emph{{Primordial physics from large-scale structure beyond the
  power spectrum}},  \href{https://arxiv.org/abs/1802.06762}{{\tt 1802.06762}}.

\bibitem{PlanckCosmo}
{Planck Collaboration}, N.~{Aghanim}, Y.~{Akrami}, M.~{Ashdown}, J.~{Aumont},
  C.~{Baccigalupi} et~al., \emph{{Planck 2018 results. VI. Cosmological
  parameters}}, {\emph{ArXiv e-prints} (July, 2018) },
  [\href{https://arxiv.org/abs/1807.06209}{{\tt 1807.06209}}].

\bibitem{BiagettiRev}
M.~Biagetti, \emph{{The Hunt for Primordial Interactions in the Large Scale
  Structures of the Universe}},  \href{https://arxiv.org/abs/1906.12244}{{\tt
  1906.12244}}.

\bibitem{GiannantonioPorciani}
T.~{Giannantonio} and C.~{Porciani}, \emph{{Structure formation from
  non-Gaussian initial conditions: Multivariate biasing, statistics, and
  comparison with N-body simulations}},
  \href{http://dx.doi.org/10.1103/PhysRevD.81.063530}{\emph{\prd} {\bf 81}
  (Mar., 2010) 063530}, [\href{https://arxiv.org/abs/0911.0017}{{\tt
  0911.0017}}].

\bibitem{Assassi:2015fma}
V.~Assassi, D.~Baumann and F.~Schmidt, \emph{{Galaxy Bias and Primordial
  Non-Gaussianity}},
  \href{http://dx.doi.org/10.1088/1475-7516/2015/12/043}{\emph{JCAP} {\bf 12}
  (2015) 043}, [\href{https://arxiv.org/abs/1510.03723}{{\tt 1510.03723}}].

\bibitem{Desjacques:2016}
V.~Desjacques, D.~Jeong and F.~Schmidt, \emph{{Large-Scale Galaxy Bias}},
  \href{http://dx.doi.org/10.1016/j.physrep.2017.12.002}{\emph{Phys. Rept.}
  {\bf 733} (2018) 1--193}, [\href{https://arxiv.org/abs/1611.09787}{{\tt
  1611.09787}}].

\bibitem{Wagner:2014aka}
C.~Wagner, F.~Schmidt, C.-T. Chiang and E.~Komatsu, \emph{{Separate Universe
  Simulations}}, \href{http://dx.doi.org/10.1093/mnrasl/slu187}{\emph{Mon. Not.
  Roy. Astron. Soc.} {\bf 448} (2015) L11--L15},
  [\href{https://arxiv.org/abs/1409.6294}{{\tt 1409.6294}}].

\bibitem{Chiang:2017jnm}
C.-T. Chiang, \emph{{Halo squeezed-limit bispectrum with primordial
  non-Gaussianity: A power spectrum response approach}},
  \href{http://dx.doi.org/10.1103/PhysRevD.95.123517}{\emph{Phys. Rev. D} {\bf
  95} (2017) 123517}, [\href{https://arxiv.org/abs/1701.03374}{{\tt
  1701.03374}}].

\bibitem{dePutter:2018jqk}
R.~de~Putter, \emph{{Primordial physics from large-scale structure beyond the
  power spectrum}},  \href{https://arxiv.org/abs/1802.06762}{{\tt 1802.06762}}.

\bibitem{Chan:2012jj}
K.~C. Chan, R.~Scoccimarro and R.~K. Sheth, \emph{{Gravity and Large-Scale
  Non-local Bias}},
  \href{http://dx.doi.org/10.1103/PhysRevD.85.083509}{\emph{Phys. Rev. D} {\bf
  85} (2012) 083509}, [\href{https://arxiv.org/abs/1201.3614}{{\tt
  1201.3614}}].

\bibitem{Baldauf:2012hs}
T.~Baldauf, U.~Seljak, V.~Desjacques and P.~McDonald, \emph{{Evidence for
  Quadratic Tidal Tensor Bias from the Halo Bispectrum}},
  \href{http://dx.doi.org/10.1103/PhysRevD.86.083540}{\emph{Phys. Rev. D} {\bf
  86} (2012) 083540}, [\href{https://arxiv.org/abs/1201.4827}{{\tt
  1201.4827}}].

\bibitem{CastorinaWhite2018a}
E.~{Castorina} and M.~{White}, \emph{{Beyond the plane-parallel approximation
  for redshift surveys}},
  \href{http://dx.doi.org/10.1093/mnras/sty410}{\emph{\mnras} {\bf 476} (June,
  2018) 4403--4417}, [\href{https://arxiv.org/abs/1709.09730}{{\tt
  1709.09730}}].

\bibitem{CastorinaWhite2018b}
E.~{Castorina} and M.~{White}, \emph{{The Zeldovich approximation and
  wide-angle redshift-space distortions}},
  \href{http://dx.doi.org/10.1093/mnras/sty1437}{\emph{\mnras} (June, 2018) },
  [\href{https://arxiv.org/abs/1803.08185}{{\tt 1803.08185}}].

\bibitem{BeutlerCastorina}
F.~{Beutler}, E.~{Castorina} and P.~{Zhang}, \emph{{Interpreting measurements
  of the anisotropic galaxy power spectrum}}, {\emph{ArXiv e-prints} (Oct.,
  2018) }, [\href{https://arxiv.org/abs/1810.05051}{{\tt 1810.05051}}].

\bibitem{Tegmark:1997rp}
M.~Tegmark, \emph{{Measuring cosmological parameters with galaxy surveys}},
  \href{http://dx.doi.org/10.1103/PhysRevLett.79.3806}{\emph{Phys. Rev. Lett.}
  {\bf 79} (1997) 3806--3809},
  [\href{https://arxiv.org/abs/astro-ph/9706198}{{\tt astro-ph/9706198}}].

\bibitem{Meiksin:1998mu}
A.~Meiksin and M.~J. White, \emph{{The Growth of correlations in the matter
  power spectrum}},
  \href{http://dx.doi.org/10.1046/j.1365-8711.1999.02825.x}{\emph{Mon. Not.
  Roy. Astron. Soc.} {\bf 308} (1999) 1179},
  [\href{https://arxiv.org/abs/astro-ph/9812129}{{\tt astro-ph/9812129}}].

\bibitem{Ferraro:2015}
S.~{Ferraro} and K.~M. {Smith}, \emph{{Using large scale structure to measure
  f$_{NL}$ , g$_{NL}$ and {$\tau$}$_{NL}$}},
  \href{http://dx.doi.org/10.1103/PhysRevD.91.043506}{\emph{\prd} {\bf 91}
  (Feb., 2015) 043506}, [\href{https://arxiv.org/abs/1408.3126}{{\tt
  1408.3126}}].

\bibitem{deMattia:2019vdg}
A.~de~Mattia and V.~Ruhlmann-Kleider, \emph{{Integral constraints in
  spectroscopic surveys}},
  \href{http://dx.doi.org/10.1088/1475-7516/2019/08/036}{\emph{JCAP} {\bf 08}
  (2019) 036}, [\href{https://arxiv.org/abs/1904.08851}{{\tt 1904.08851}}].

\end{thebibliography}\endgroup
\end{document}